\documentclass[12pt]{article}
\usepackage[utf8]{inputenc}

\usepackage{amssymb,amsmath,amsthm,amsfonts,eurosym,geometry,ulem,graphicx,caption,color,setspace,sectsty,comment,footmisc,caption,natbib,pdflscape,array,hyperref,subcaption,pgfplots,appendix}

\DeclareMathOperator*{\argmax}{arg\,max}

\theoremstyle{plain}

\newtheorem{result}{Result}

\theoremstyle{definition}

\begin{document}

\begin{titlepage}
\title{Artificial Intelligence and Auction Design\thanks{We thank seminar participants for comments and suggestions.}}
\author{Martino Banchio\thanks{Stanford Graduate School of Business, Stanford University. Email: \href{mailto:mbanchio@stanford.edu}\texttt{mbanchio@stanford.edu}} \and Andrzej Skrzypacz\thanks{Stanford Graduate School of Business, Stanford University. Email: \href{mailto:skrz@stanford.edu}\texttt{skrz@stanford.edu}}}
\date{\today}
\maketitle
\begin{abstract}
\noindent Motivated by online advertising auctions, we study auction design in repeated auctions played by simple Artificial Intelligence algorithms (Q-learning). We find that first-price auctions with no additional feedback lead to tacit-collusive outcomes (bids lower than values), while second-price auctions do not. We show that the difference is driven by the incentive in first-price auctions to outbid opponents by just one bid increment. This facilitates re-coordination on low bids after a phase of experimentation. We also show that providing information about lowest bid to win, as introduced by Google at the time of switch to first-price auctions, increases competitiveness of auctions.\\
\vspace{0in}
\vspace{0in}\\

\bigskip
\end{abstract}
\setcounter{page}{0}
\thispagestyle{empty}
\end{titlepage}
\pagebreak \newpage

\maketitle

\section{Introduction.}
In this paper we revisit a classic question: how does auction design affect revenues and bidder behavior? We shed a new light on this question by analyzing auctions where bidders use simple artificial intelligence algorithms to determine their bids (rather than using the Nash equilibrium paradigm). 
Our main result is that when bidders use these algorithms, auction format and other design choices can have a first-order effect on revenues and bidder payoffs. 

In particular, we present two findings. First, we observe that revenues can be significantly lower in first-price auctions than in second-price auctions. We show this in a very simple setup: with two bidders and constant symmetric values. The algorithms we consider are simple Q-learning algorithms that keep track of the history of past auctions in a very reduced form and are not explicitly designed to discover dynamic reward/punishment strategies. 
Second, we show that a simple practical auction design choice --- revealing bidders the winning bid after the auction --- can make the first-price auctions much more competitive.\footnote{When Google switched in 2019 the Google Ad Manager auctions for display advertising from the second-price to the first-price format, they simultaneously started providing such information to all buyers bidding in that auction: “Buyers will receive the minimum bid price to win after the auction closes” - see \url{https://blog.google/products/admanager/rolling-out-first-price-auctions-google-ad-manager-partners/} .}  In fact, providing such information can bring bids and revenues to the competitive level predicted by a one-shot Nash equilibrium (and to the level we observe in second-price auctions).

We are interested in studying how auction design affects play by AI algorithms for two reasons. First, we think that our analysis is quite applicable to online advertising auctions, where bidders for some types of impressions/clicks compete in many thousands of auctions a day.\footnote{The number varies greatly with the website and the target audience. To get a sense of the volumes, the New York Times website is estimated to be visited 339M times in November 2021 and visitors viewed on average 2.28 pages per visit - see \url{https://www.similarweb.com/website/nytimes.com}. With even only one ad per page, that is over 25M ad impressions sold a day. Estimates of the number of daily searches (and hence opportunities to run keyword auctions) on Google’s search engine are in billions a day. Clearly, any given advertiser is interested in only a fraction of those advertising opportunities, yet we still expect that many advertisers participate in thousands of auctions a day.} 
Online ad auctions usually happen in a fraction of a second. It is common for bidders to rely on bidding tools either provided by the auctioneers, developed in-house or by third parties.\footnote{There are many such examples: Smart Bidding at Google (\url{https://support.google.com/google-ads/answer/7065882?hl=en}) is described as “a subset of automated bid strategies that use machine learning to optimize for conversions or conversion value in each and every auction.” An Example of a third-party algorithm is scibids.com that states “We build customizable AI that dramatically improves Paid Media ROI.”}. As such algorithms grow in popularity, it is a natural question to ask how play in such auctions would evolve when multiple competitors use similar tools, each optimizing on behalf of its owner.

Second, a collection of recent papers on algorithmic pricing expressed concern that algorithms used for pricing could facilitate (tacit) collusion (see the related literature discussion). The general theme is that sophisticated algorithms can learn to play dynamic strategies, offering rewards to their competitors for tacit collusion and punishing them for competitive behavior --- playing a low-revenue equilibrium of an infinitely repeated game instead of the high-revenue equilibrium of the static game. Such behavior by algorithms can be facilitated by their ability to continuously monitor each other and react (with punishments) at much higher speeds than humans can. Our theory of repeated games suggests that improved monitoring and speed of reaction make tacit-collusive equilibria more stable. We are interested in understanding to what extent the repeated games intuitions apply to simple AI algorithms and whether auction design choices can affect those forces.

The simple AI algorithm we use in this paper is Q-learning, one of the most popular algorithms for agnostic learning. We consider the simplest Q-learning algorithm that works roughly as follows: for a finite set of available bids, a bidder keeps track of a Q-vector, that is its current estimate of the value of taking any action (the Q-vector has one entry for each potential bid). The Q-vector tries to estimate both current and future payoff consequences of choosing any particular bid today. The algorithm chooses with probability $1-\epsilon$ in any given period a bid that corresponds to the highest expected entry in Q. With probability $\epsilon$ it takes any other action. As the player observes its current payoff, he updates its Q estimate, putting a higher weight on recent data. In our main specification, epsilon decays over time, capturing the tradeoff between exploration and exploitation.\footnote{We also show some results in case epsilon remains positive even in the long run and they are the same as our main finding. A more complex Q-learning algorithm would include additional states - for example, allowing the estimate of Q to depend on recent history, like who won the last auction. Such more sophisticated algorithms are of interest. We do not discuss them in this paper since they are much harder to analyze and because they are by design guiding bidders towards tacit collusion. One of the things we are interested in this paper is understanding whether tacit collusion is a concern even with simple AI algorithms.}

The Q-learning algorithm has been shown to have great success in finding optimal strategies in single-person decision making problems — see for example \citet{Watkins1992}. In our setup, the algorithm performs very well at finding the best response strategy to any fixed strategy of its opponents. Additionally, Q-learning algorithms are the basic building blocks of more sophisticated AI algorithms, therefore we think understanding the dynamics of bidding by Q-learning agents is going to shed light on likely consequences of auction design in more complex environments. 

Our first main finding is shown in Figure \ref{fig:base}, representing the distribution of long-run behavior for our bidding agents over 1,000 experiments that each involve 1,000,000 auctions. The results are stark: bidders in the repeated second-price auctions converge to bidding according to the static Nash equilibrium prediction and the revenues to the auctioneer are high. In contrast, in the first-price auctions bidders converge over time to much lower bids (the average bid is 0.24) --- this seems to be reminiscent of tacit collusion.\footnote{ These results are robust to not letting the experimentation rate $\epsilon$ decay to zero --- see Figure \ref{fig:exp}. }

Our second main finding is shown in Figure \ref{fig:synchronous}: when in the first-price auction bidders observe not only whether they won, but also what was the lowest bid to guarantee winning, they can update the Q vector not only for the current bid, but for all counterfactual bids (so-called synchronous updating). It turns out that with this additional information, even in the first-price auction bidders converge over time to highly competitive bids. 

To understand the economic forces behind the first result, we notice that there are multiple differences between the first-price and second-price auction formats that can contribute to the difference in outcomes, despite the theory of repeated games suggesting that tacit collusion should be approximately equally easy/hard in those two games. In particular, we point out that:
\begin{enumerate}
    \item The second-price one-shot auction has a dominant bidding strategy while the first-price auction does not, and the algorithm is better at finding dominant strategies.
    \item When bidders try to coordinate on a low bid, then in the second-price auction all higher bids are profitable (in the short run) while in the first-price auction only bids close to the current bid are profitable. For example, if bidders coordinate on bids (0.2, 0.2), then any bid between 0.2 and 1 is profitable in the SPA while only bids  between 0.2 and 0.6 are profitable in the FPA. Since our algorithms discover profitable deviations via random experimentation, it may be harder for them to find those in FPA, some tries to a higher bid (for example, 0.8) can result in lower profits than current tacitly collusive bid, and hence bidders may stop experimentation.
    \item Because there are so many more potentially profitable deviations, it is possible that it is harder for the algorithms to coordinate/reach the tacit-collusive outcome in the second-price auctions than in first-price auctions, so the difference comes from the algorithms' initialization. However, with an initialization on low bids, the long-term conduct could be the same
    \item The two auctions are different in terms of how the algorithms behave when they mis-coordinate at different bids. Namely, when bidder one outbids bidder two, in both formats bidder two will learn that their current strategy is not profitable and will start exploring other bids. The difference is in how bidder one (the higher bidder) behaves in these two formats. In the second-price auction, the payoffs of that bidder do not depend on his current bid, hence there is no first-order force to push him to higher or lower bids. On the other hand, in the first-price auction, a winning bidder has incentives to win by as little as possible. So if the bidders start with bids (0.2, 0.2), the deviating algorithm is going to learn that against a constant opponent bid the optimal deviation is just by one bid increment (by 0.05 in our simulations). As a result, when the bidders submit equal bids, they on average hit lower bids in the first-price auction than in the second-price auction.
\end{enumerate}

We designed additional experiments/simulations to tell these possible explanations apart. It turns out that the main reason for the difference is the last one. This force is a fundamental difference between the first-price and second-price auctions. It also helps us understand why the tacit collusion seems to lead to an average revenue 0.24 per auction and not less. When bidders try to coordinate on very low bids, experimentation pushes them only up not down, and they are very unlikely to return to very low bids.

To understand the forces behind the second result, it is helpful to think of the incentives to re-coordinate on lower bids. Imagine that both bidders coordinate on a bid of $0.3$, until bidder one deviates to $0.4$. Bidder one will experience an immediate boost in his estimate of the new bid's value, while bidder two will surely realize that $0.3$ is no longer a good bid. After some experimentation, suppose the two coordinate on $0.4$. If the two bidders can only update the current bid, the bid $0.3$ remains biased: soon enough they will discover that coordinating on $0.4$ is worse than coordinating on $0.3$ and attempt to move back. If they update synchronously instead, using the counterfactual return from other bids, the estimate of the value of $0.3$ will drop dramatically: the return in hindsight is always zero. Synchronicity in some sense leads to shortsightedness: the counterfactual measure used does not take into account future re-coordinations.

We finish the paper by considering several extensions. We analyze the effect of reserve prices on bidder behavior and compare it with the introduction of a competitive fringe. We analyze the game with more competition - either with 3 strategic bidders or two bidders and a competitive fringe (as expected, competition increases revenues in the first-price auctions, but it does not eliminate the tacit-collusive outcomes).

\subsection{Related Literature.}

Algorithmic collusion has recently sparked some interest in the Economics community. The pioneering work of \citet{Calvano2020} examines collusion in a price-setting oligopoly, and suggests that algorithms  keeping track of past prices adopt collusive strategies typical of implicit cartels. They too study Q-learning as a workhorse Artificial Intelligence algorithm. \citet{Klein2021} also studies collusion and Q-learning in a pricing game, but in that setting price offers are alternated. In a similar pricing model, \citet{Asker2022} study the effect of algorithm design on collusion. Similar to their paper, we find that synchronous algorithms are less likely to converge on collusive outcomes. The strength of such finding in our setting is supported by Google's auction design choice, as described in the introduction.  \citet{Hansen2021} simulate a different algorithm from the bandit literature, and show how its misspecified implementation similarly results in collusion. 

Empirical work on algorithmic pricing has also been flourishing. \citet{Brown2021} study the effect of high-frequency price adjustments on competition between online retailers. They find that higher frequency may lead to price dispersion and increase overall price levels.  \citet{Musolff2021} finds that online retailers' prices follow a Hedgeworth's cycles behavior through a price-resetting mechanic that decreases competition. In \citet{Assad2021} the authors document a rise in margins of retail gasoline sellers from adoption of automated pricing algorithms. Some recent work analyzes learning in auctions, but most of the literature is concerned with the auctioneer's side (\citet{Milgrom2019}) with the exception of \citet{Nedelec2019}: they analyze a strategic bidder's objective and approach learning through gradient descent. 

The pioneering work on Q-learning by \citet{Watkins1989} and \citet{Watkins1992} revived a literature on reinforcement learning which has been explored briefly also in economics (\citet{Erev1998}, \citet{Erev1999}). The theory surrounding multi-agent learning has had great success in practical applications, but there is no consensus among computer scientists on a leading paradigm. For a recent overview, we refer the reader to \citet{Zhang2021}. The theoretical studies on multi-agent reinforcement learning have led to connections with the evolutionary game theory literature (see \citet{Bloembergen2015}), mostly for specific algorithms and rules. Recently, \citet{Banchio2022b} propose a continuous-time approximation for algorithmic systems which allows for characterization of equilibria. Our setting is too high-dimensional to apply such techniques, but we are able to test our hypotheses experimentally and build intuitions from their simple setup. 

The literature on Learning in Games has analyzed systems of learners from various angles. The workhorse model of Fictitious Play by \citet{Brown1951} has been studied thoroughly and its properties are well understood (see \citet{Fudenberg1998} for a thorough review), but its practical adoption has been long shunned. The simple reinforcement learning models of \citet{Erev1998} and \citet{Borgers1997} produce interesting predictions with simple algorithms. However, most of the literature is concerned with learning as a foundation for Nash equilibrium and equilibrium concepts as a whole. In this work we are interested in the equilibrium behavior of learning systems instead, where the learning system is taken as given and equilibria are hard to characterize. As we show, in some auction formats the algorithms do not converge to the Nash equilibrium of the static game. This means that in a play where experimentation does not die out, they do not converge to any one action profile. Instead they end up in stochastic cycles, with long-term average bids substantially lower than in the Nash equilibrium of the static game.

The final literature our paper is related to is theoretical literature on collusion in auctions. \citet{Mcafee1992} discuss collusive schemes by strong and weak cartels, defined as those that use and do not use side payments. They discuss the properties of the bid rotation schemes and under what conditions bidders can benefit from them. \citet{Skrzypacz2004} study tacit collusion in repeated games where the bidders observe publicly only the identity of the winner (see also \citet{Aoyagi2003} for further analysis of bid rotation in repeated auctions). \citet{athey2001} study tacit collusion when players observe all bids. \citet{Marshall2009} show how the extensive form of auction formats may inhibit or facilitate tacit collusion. Our contribution is to show the interaction between auction design and level of competition while assuming strategies are chosen by the algorithms rather than they are dictated by rationality assumptions.

\section{The Model.}\label{sec:model}
Two bidders participate in a sequence of auctions.\footnote{Some of our simulations consider more than two bidders; we explain those extensions later.} In every period $t \in \{1,\dots,\infty\}$ an auctioneer runs an auction to allocate a single non-divisible object to one of the bidders. Both bidders value the object at $v_i = 1$ and the value is constant over time. 

We consider a family of auction formats parameterized by $\alpha \in [1,2]$. In an $\alpha-$auction the highest bidder wins and pays a convex combination of the winning and the losing bid. The weight on the losing bid is $\alpha-1$, and the weight on the winning bid is $2-\alpha$. We focus on the two extreme cases: the first-price auction (FPA, $\alpha=1$) and the second-price auction (SPA, $\alpha=2$). 

The payoff of the winner of period $t$ auction is $\pi_t = 1-p_t$ where $p_t$ is the price determined by the mechanism chosen by the auctioneer. The losing bidder gets a payoff of $0$. Bidders maximize the expected sum of discounted per-period payoffs with a discount factor $\gamma \in (0,1)$. 

In the auction, bidders choose from a finite grid of prices. Each bidder has access to a set of equidistant bids $[b_1, \dots, b_m]$ where $b_i = \frac{i}{m+1}$. This assumption allows us to work with simpler learning algorithms. It is also representative of auctions typically allowing bidders to submit bids expressed in dollars and cents. Some online auctions restrict the bids even further.\footnote{For example, in Capterra.com auctions ''Bids start at \$2  per click and can be increased in \$ 0.25 increments.'' See https://blog.capterra.com/what-is-ppc/.} Note that we restrict attention to bids smaller than the bidder values.\footnote{We also assume that the highest bid is strictly less than value so that in the static Nash equilibrium, bidders play strict best responses. When the highest available bid is $1$, the payoffs are zero in the static Nash equilibrium, and bidders are indifferent between following the equilibrium strategy and any deviation to a lower bid. To ensure this indifference does not drive our results, we keep a small wedge between the highest bid and the value. An additional benefit of our assumptions about the grid is that the equilibrium of the second-price auction is unique (while it is not unique when bids can be arbitrary).}

\subsection{Nash Equilibria of the Auctions.}

This paper aims to study how auction design (for example, a choice between the first-price and second-price auction) affects the outcomes if the players use simple artificial intelligence algorithms to choose their bids. 
To put our results in perspective, we first discuss Nash equilibria of the auctions from the repeated and static perspectives. 

Given our very simple environment, for all $\alpha$, the one-shot game has a unique Nash equilibrium, with both players bidding $b_m$ (the largest bid bellow value). In equilibrium, bidders have strict incentives to follow the equilibrium strategy. In these static equilibria,  auctioneer revenues are independent of the format.

If the discount factor is sufficiently high, the repeated game has many other equilibria. The set of equilibria depends on the information provided to the bidders after every auction. Do they only observe whether they won and their price? Or do they observe both bids? Alternatively, do they observe something else?
The analysis of the equilibria of the repeated auctions is simpler when the bidders observe both bids after the auction. That makes it a game with perfect public monitoring. If bidders only observe their bids and whether they won, this becomes a game with imperfect public monitoring. It is often the choice of the auctioneer how much information to reveal to the players. We analyze the consequences of such a design choice in Section 4.\footnote{For example, in FPA, the auctioneer can choose to reveal or hide the losing bid from the winner and/or reveal the winning bid to the loser. As we mentioned in the Introduction, in a recent switch from SPA to FPA, Google decided to reveal both the winning bid to losers and the highest losing bid to the winner).} 

To keep this discussion short, we review only some of the equilibria. In Appendix \ref{app:equilibria} we first discuss strongly symmetric equilibria when the bidders observe both bids. Then we consider Bid Rotation equilibria that require only that the bidders observe the identity of the winner\footnote{For an analysis of Perfect Public Equilibria of the repeated FPA and SPA with private values and bidders publicly observing only the identity of the winner see \citet{Skrzypacz2004}}. The takeaway from the repeated games literature is that, with perfect monitoring, strongly symmetric collusive equilibria are easier to sustain in the FPA than in the SPA, although the difference is minor.  Instead, if players only publicly observe the winner's identity, it is possible to sustain tacitly collusive outcomes even via simple public perfect equilibria, through bid rotation schemes. Moreover, the analysis suggests that these equilibria should be easier to sustain in SPA than in FPA. 
The main takeaway from our results is that the propensity of simple algorithms to reach tacitly collusive outcomes depends on economic forces beyond the intuitions from the analysis of the repeated games.

\section{Q-Learning.} 

First proposed by \citet{Watkins1989}, Q-learning algorithms are the main building blocks of the reinforcement learning paradigm. Actions found to be more profitable are more likely to be taken in the future: each result reinforces the agent's understanding of the environment.
Formally, in each period two agents choose a bid $b^i_t \in B = \{b_1,\dots,b_m\}$. The agents earns a stochastic reward $r_t$ distributed according to $F(r_t|b^i_t,b^{-i}_t)$. 
We will work with a simple tabular environment with finite action set $B$ not just for simplicity but for interpretability: the model requires few hyperparameters with clear economic relevance, while the neural networks necessary to implement more complex Q-learning-based approaches do not easily lend themselves to interpretation. A convenient simplification is absence of states for the algorithm. While some papers design the learners to keep track of past actions, in the original Q-learning formulation states are Markovian parameters of the environment. In this sense, our environment is time-independent, and the algorithms do not need any additional information about play.\footnote{Q-learning is not misspecified here: rewards in each period depend only on the actions of the agent in that period. The ability to recall past actions serves as a monitoring technology, but does not have direct payoff implications. Contrast this with Markov games where the distribution of rewards conditions on the current state of the environment.} Additionally, some of the environments discussed in the introduction do not provide enough information to the algorithms for appropriate memory representation.

Each agent maximizes the discounted sum of rewards $\mathbb{E}\Big[\sum_t^\infty \gamma^t r_t\Big]$ where $\gamma <1$ is the discount factor. Instead of considering the value of Dynamic Programming, Q-learning estimates the action-value function:
\[
Q(a) =  \mathbb{E}[r|b^i,b^{-i}] + \gamma \mathbb{E}[\max_{b'} Q(b')]
\]
Notice that the optimal value is simply $V = \max_b Q(b)$. If the agent learns the Q-function, he can play the optimal strategy. The algorithms is simple: starting from an arbitrary initial action-value function, after choosing an action $a_t$  update the Q-function as follows:
\[
Q_{t+1}(b_t) =  (1-\alpha)Q_{t}(b_t) + \alpha\Big[r_t + \gamma \max_b Q_t(b)\Big]
\]

This particular form of learning is asynchronous: only the state-action pair visited in a particular period is updated, while the rest of the Q-function remains constant.
The hyperparameter $\alpha$ is called learning rate. Its task is to discipline the speed of learning, but also for how long past experience is retained in today's estimate of action-values. The updating procedure is a long-run average, and in this sense the parameter $\alpha$ is the counterpart of the discount factor: $\gamma$ determines the importance given to the future, while $\alpha$ specifies how quickly the algorithm forgets about the past. 

\citet{Watkins1992} prove that Q-learning converges to the optimal policy in a Markov Decision Problem (MDP) for a single agent. However, no such guarantee exists for general multi-agent Q-learning. Difficulties arise from the loss of stationarity: each agent faces an unpredictable, ever-changing environment. The reward distribution depends on the opponents actions as well. One approach to multi-agent Q-learning considers opponents' past actions as part of the state, but essentially ignoring the endogeneity of transition and reward probabilities. While the Markov property is clearly not satisfied, various experiments in the literature find independent Q-learning to perform well in these settings, as is the case in our repeated auction. Additionally, opponent-aware algorithms would require more information about each opponent's design and behaviour, whereas the independent design approach retains the model-free philosophy of the reinforcement learning paradigm.

\paragraph{Experimentation.}
The Q-learning procedure specifies an update policy for every action taken, but it does not specify a choice of action directly.  In the algorithm proposed by Watkins, agents take actions uniformly at random: repetition guarantees that each sequence of actions will be taken sufficiently many times, and Q-learning will eventually visit every state and learn its value. This approach is limited in multiple ways, and particularly it fails to account for the tradeoff between exploration and exploitation: after an initial exploration period it would be reasonable to reap profits from actions that have been consistently outperforming.
We focus on a rule known as $\epsilon$-greedy: the agent takes the action that maximizes the Q-function with probability $1-\epsilon$, and takes an action uniformly at random with probability $\epsilon$. The exploration probability will take the form $\epsilon = \varepsilon e^{-\beta t}$, where $\beta$ further regulates the exploration-exploitation tradeoff as time progresses.

Another popular exploration paradigm, optimistic Q-learning, has shown some degree of success. Optimistic Q-learners are purely greedy: they always take the estimate of the optimal action at that point in time. However, the Q-function is initialized unusually high: for every state-action pair, the value of $Q$ is larger than the maximum payoff it could ever be achieved. The purpose of such an initialization is to ensure that experimentation will be pervasive: the algorithm won't stop experimenting until all of the $Q(a)$ values will have sufficiently decreased.\footnote{\citet{EvenDar2002} prove that optimistic tabular Q-learning converges to the correct optimal policy in MDPs.} In our multi-agent setting, the advantage offered by optimism is a phase of intense experimentation at the beginning, which improves convergence.

\section{Results.}
In most of our simulations, we simulate $2$ independent Q-learning algorithms bidding in the auctions. The grid of allowable bids includes 19 bid levels from $0.05$ to $0.95$. Unless otherwise stated, each experiment is repeated $1000$ times, and we terminate each run after 1 million periods. The algorithms have converged if the strategy does not change for the last 1,000 iterations, that is, if for each player $\argmax Q_i(a)$ stays constant.  We discard simulations that do not converge\footnote{Note that in the baseline specification, 1 million periods is enough to obtain convergence of nearly all experiments. However, to ensure robustness, we also replicate the experiments with 10 and 100 million periods, with nearly identical results.}. In the baseline specification, we adopt an $\epsilon$-greedy exploration policy with $\varepsilon = 0.025$, $\beta = 0.0002$, $\gamma = 0.99$, and $\alpha = 0.05$, with an optimistic initialization.

\begin{figure}
\centering
\begin{subfigure}{.5\textwidth}
  \centering
  \includegraphics[width=83mm]{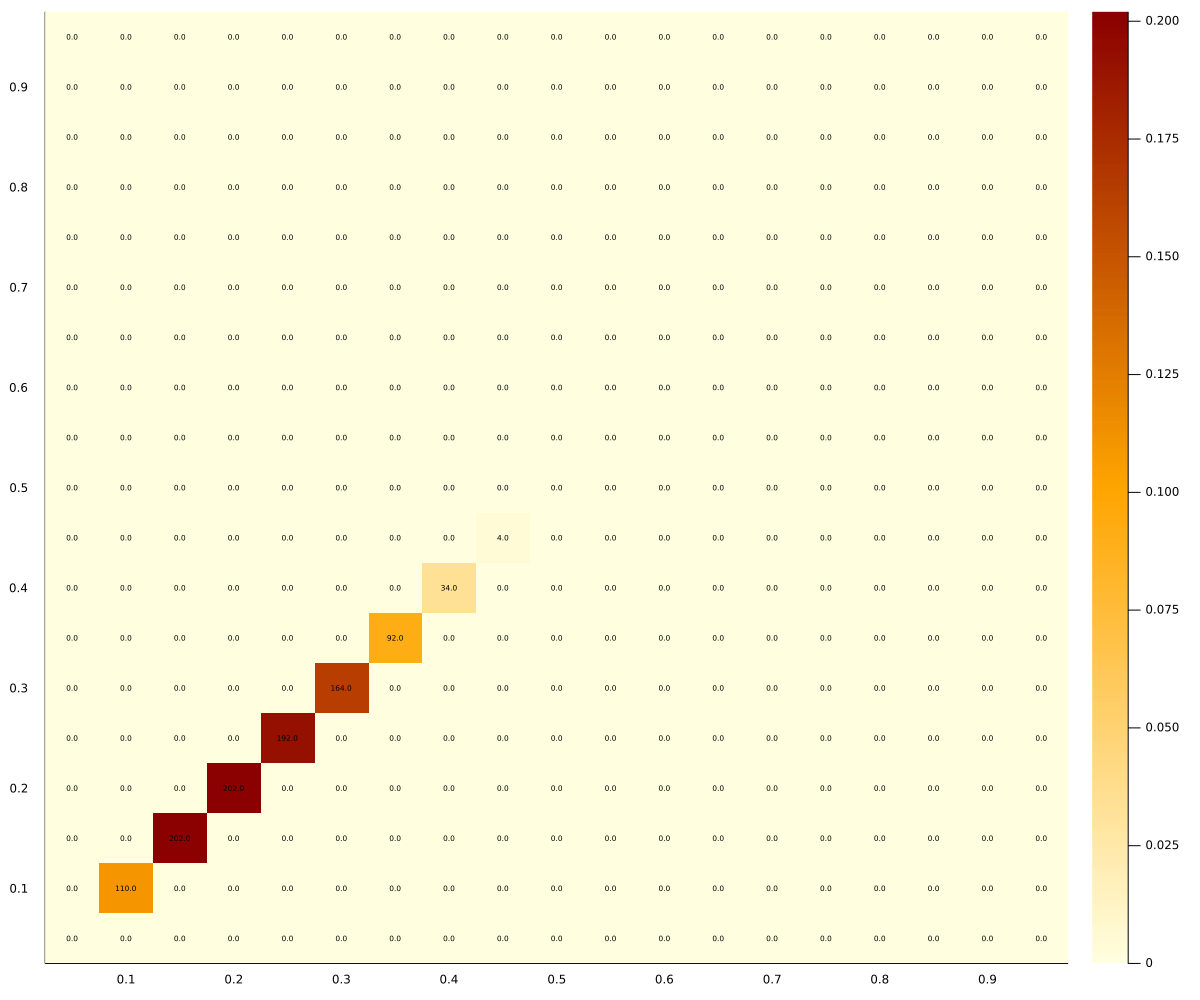}
  \caption{FPA bids}
  \label{fig:baseFPA}
\end{subfigure}%
\begin{subfigure}{.5\textwidth}
  \centering
  \includegraphics[width=83mm]{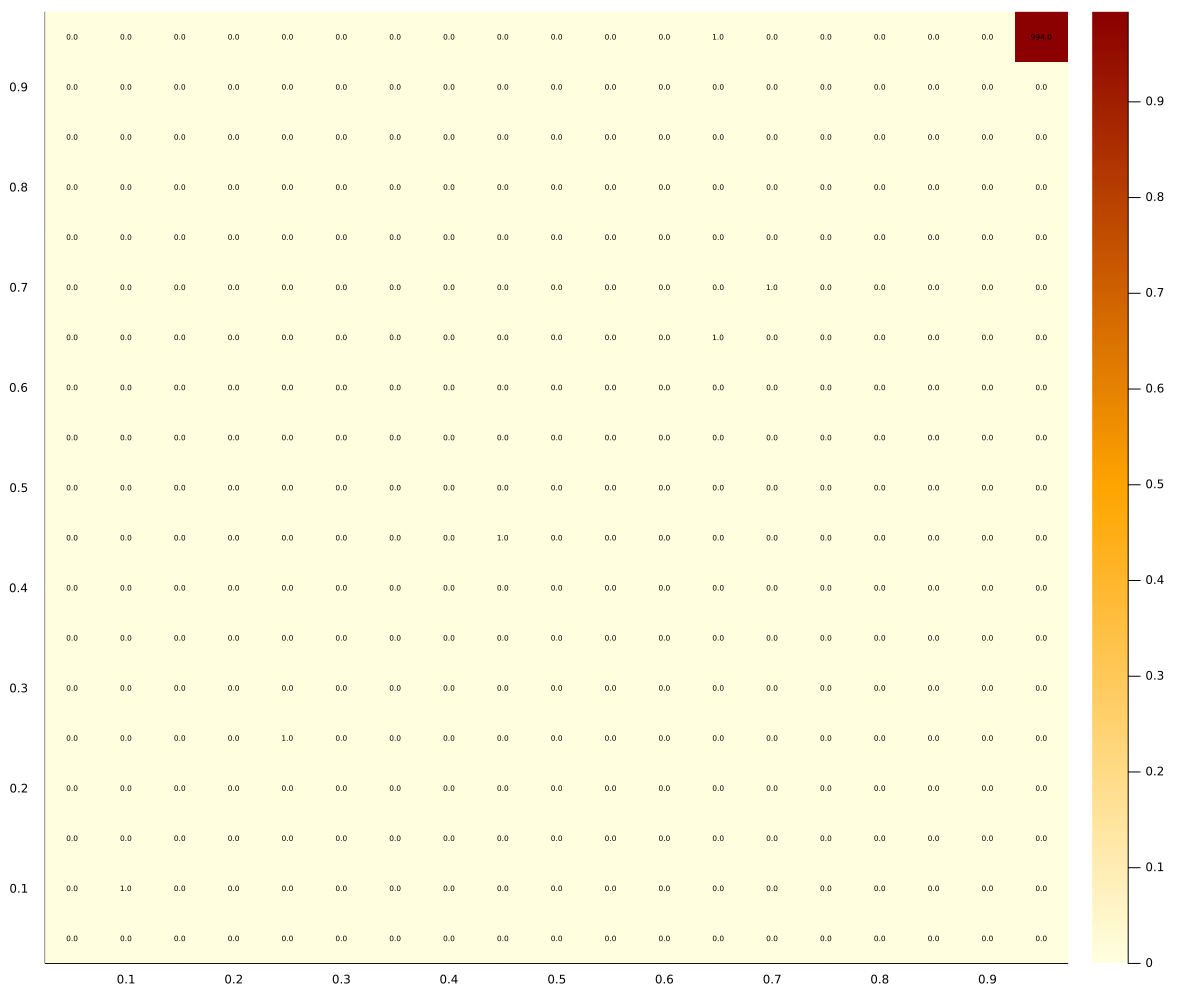}
  \caption{SPA bids}
  \label{fig:baseSPA}
\end{subfigure}
\caption{The heatmaps above show the frequencies of each pair of strategies at convergence. The bids of the two algorithms are ordered along the $x$- and $y$-axis. }
\label{fig:base}
\end{figure}

In our first set of results We compare bidding outcomes under the first-price auction (FPA) and second-price auction (SPA) formats. The outcomes of these experiments are reported in Figure \ref{fig:base}.

\begin{result}\label{Result1}
\item Our algorithms converge to the static Nash equilibrium in the second-price auction. 
\item They converge to much lower bids in the first-price auction.
\item There is large dispersion of outcomes at convergence in FPA, and no dispersion in SPA.
\item The average (across simulations, at convergence) revenue per auction in SPA is $0.95$ and in FPA it is $0.24$.
\end{result}

One thing to notice in Figure \ref{fig:base} is that the algorithms do not converge on pairs of bids outside of the diagonal. This is intuitive since if they did, the losing algorithm would learn that their strategy gives payoff $0$ and would over time find out that deviating to bid $0.95$ would be profitable (a contradiction).

The difference between the outcomes for the two formats is rather striking: the two bidders converge to the static equilibrium in the case of SPA. But they ``collude'' on low bids in FPA. In particular, the average revenue of the auctioneer when using SPA is $0.95$ (the highest possible bid), while the average revenue under FPA is $0.24$. 

Note that the equilibrium theory of repeated games that we discussed earlier is not good at predicting this outcome. As we explained, tacit collusion based on strongly symmetric equilibria is sustainable under both formats for similar assumptions, and tacit collusion based on asymmetric equilibria is much easier to sustain in the second-price auctions. A different view could be that since our algorithms do not explicitly keep track of history (they are not designed to learn conditional strategies like the bid rotation scheme), Nash equilibrium theory would predict that collusion should not be possible in either of the formats.

One may worry that the decay of experimentation drives our results. Our algorithms may be getting stuck at low bids in FPA because they stop deviating by the time they learn that there is a profitable deviation. To check that our Result \ref{Result1} is robust to other processes for experimentation, we also ran the auctions for $100$ million iterations, while keeping the experimentation parameter constant at $\varepsilon = 0.001$. The results are presented in Figure \ref{fig:exp} that counts the number of times out of the 100 million rounds the different pairs of bids have been played. Since these algorithms never stop experimenting, they continue to visit all pairs of bids. Yet, consistent with our findings in Result \ref{Result1}, a clear pattern appears: the algorithms spend most of the time in SPA at bids $(0.95,0.95)$ and much lower bids in FPA.

Interestingly, in FPA they do not spend all the time at one pair of low bids but move across them. This is consistent with the algorithms being good at finding best responses - if an opponent converged to a constant bid (in most periods), an algorithm in FPA would learn to bid just one bid increment more. So, the only candidate for convergence to constant bids is the static Nash equilibrium. Instead, the algorithms end up in a cycle, moving between several pairs of low bids.

Our next step is to formulate multiple hypothesis for the main economic forces that cause the differences and then design additional simulations to test them. We then try to also provide economic intuition for why in the first-price auction despite the algorithms seemingly learning to tacitly collude, they converge to bids far away from perfect collusion at $b_i=0.05.$

\subsection{Hypotheses.}
We now formulate and analyze a few possible explanations for the observed play by algorithms in the two formats. Our methodology is to design experiments/simulations for each hypothesis to test if it seems to be one of the key forces responsible for those results. 

\paragraph{Dominant strategy.} One difference between FPA and SPA pointed out in the literature is the ``simplicity'' of the latter. The equilibrium strategy in the SPA is the unique weakly dominant strategy hence does not require conjectures about the play of the opponent. This is not true in the FPA - there the best response depends on the conjecture about the distribution of bids of the opponent. 
Moreover, from the learning literature we know that strategic simplicity (for example, the game being dominance-solvable) is often enough to guarantee convergence on Nash equilibria. 

\begin{figure}[h]
\centering
\begin{tikzpicture}
\begin{axis}[ylabel near ticks,
    xlabel near ticks,
    xlabel={$\alpha$},
    ylabel={\%}, ymax=100,ymin=0, minor y tick num = 3, area style, colormap/jet,]
\addplot+[ybar interval,mark=no] plot coordinates { (1, 100) (1.1, 100) (1.2, 100) (1.3, 100) (1.4, 86) (1.5, 41) (1.6, 15) (1.7, 2) (1.8, 0) (1.9, 0) (2, 0)};
\end{axis}
\end{tikzpicture}
\caption{Percentage of simulations that converge on a collusive outcome for different values of $\alpha$.}
\label{fig:alpha}
\end{figure}
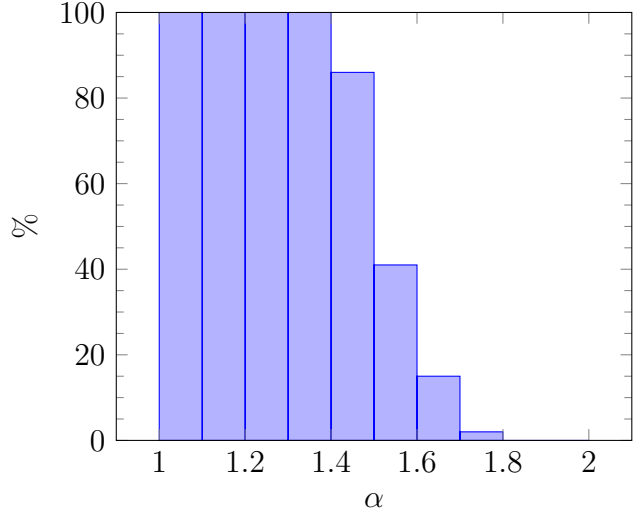

To see whether dominance per se affects the results, we run the following experiment. We let the bidders compete in a $\alpha$-price auction: the highest bidder gets the object and pays a $\alpha$-convex combination of the first- and second-highest bid. The static Nash equilibria for all of these auctions are the same - both bidders choose $b_i=0.95$. The results are summarized in Figure \ref{fig:alpha}. For low values of $\alpha$ we observe the dominant strategy equilibrium, for high values we observe collusion, and intermediate values may lead to either outcome.

\begin{result}
\item The fraction of times our algorithms converge to the static-Nash equilibrium in $\alpha$-price auctions is increasing in $\alpha$. When $\alpha$ is close to $2$, they always converge to the fully-competitive outcomes and the fraction drops gradually for lower $\alpha$.

\end{result}

Notice that, except for $\alpha = 2$, none of these formats has a dominant strategy. For all $\alpha<2$ the best response to any bid less than $0.95$ is to bid one increment above. This result allows us to conclude that strategic simplicity is not the main driver of the observed differences between FPA and SPA.

\paragraph{Few profitable deviations.} The second hypothesis is that the difference is caused by the following property of the FPA and SPA. Suppose the two bidders bid  $(0.35, 0.35)$. In a SPA any bid above $0.35$ is a profitable deviation in the immediate future, because it simply doubles the expected return for the deviating bidder. In contrast, in a FPA, only bids above $0.35$ and below $0.65$ are profitable: above $0.65$ the expected profit of the deviating bidder is lower than with the bid $0.35$. It is possible that when our algorithms randomly experiment in search for a better strategy, the have a harder time finding one when the set of immediately profitable deviations is smaller. And as a result, they may ``get stuck.''\footnote{Moreover, the set of deviation bids that are profitable is increasing in $\alpha$, so this explanation could also explain the pattern we observed in  Figure \ref{fig:alpha}.}

To test this hypothesis, we restrict the domain of experimentation of our algorithms. We constrain the algorithms to local deviations: only bids immediately above or below are admissible deviations from the currently optimal strategy. When experimentation is only local, the difference in the shape of the payoff is reduced, and particularly the difficulty in finding the profitable deviation is eliminated. Now for both SPA and FPA one out of the two possible deviations is profitable and one is not. 

We present the results in Figure \ref{fig:local} in the Appendix. The results do not change substantially: there is slightly less coordination on the dominant strategy in SPA and slightly lower bidding in FPA, but the general pattern remains.

\begin{result}
\item Modified algorithms that experiment only locally converge to approximately the same outcomes as in Result \ref{Result1}: static-Nash equilibrium bidding in SPA and much lower bids in FPA. 
\item In other words, our results are robust to changing the experimentation method of the algorithms from global to local.
\end{result}

\paragraph{Collusion is hard to discover?} Another possible explanation is that the different nature of the games makes it harder for the algorithms to discover tacit collusion in the SPA than in the FPA. We chose to use optimistic initialization of the algorithms that leads to a considerable period of exploration. It is possible that this phase, rich of uncertainty, prevents the bidders from finding a good collusion outcome in SPA. To test this hypothesis, we initialize the algorithms at a collusive outcome. Naturally, if the experimentation parameter $\epsilon = \varepsilon e^{-\beta t}$ is too low, the algorithms never leave their initialization. However, we find that with enough experimentation, the FPA remains collusive (not necessarily in the outcome it had been initialized to), while the SPA reverts back to dominant strategy.

\begin{figure}
    \centering
    \includegraphics[scale=0.65]{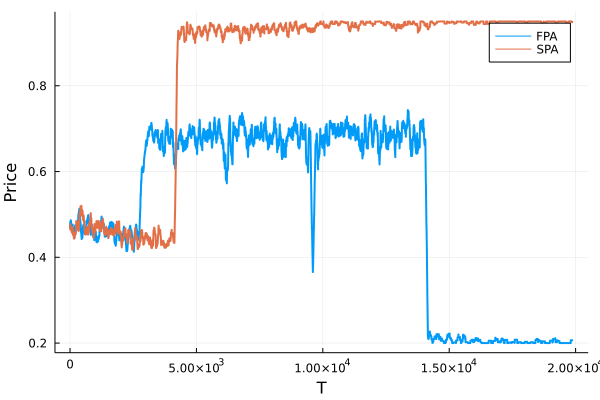}
    \caption{The picture shows the moving average of the winning bid in a single simulation for both FPA and SPA initialized with a bias towards collusion at bids of $(0.4,0.4)$. }
\label{fig:initialization}
\end{figure}

Figure \ref{fig:initialization} shows that with exploration parameter $\epsilon = 0.25 e^{-0.0002 t}$ the second-price auction overcomes its bias towards collusion, reverting back to perfect competition. The first-price auction remains collusive instead.

\paragraph{Incentives to deviate to near-by Bids.}
The final difference we explore is that in SPA when a bidder finds a profitable deviation by bidding more than the opponent, every higher bid is equally profitable. In contrast, in FPA when a bidder deviates to a higher bid, even if it is profitable (and as we pointed out before, not all higher bids are profitable in FPA), the bidder should learn that lowering their bid to just above their opponent is even better. As a result, when bidders get outside a temporary coordination at equal bids, the next time they start bidding the same amount, the bids tend to be lower in FPA than in SPA. This helps the bidders converge to a local cycle at low bids instead of getting stuck at the static Nash equilibrium. Our simulations show the most support for that economic force.

\begin{figure}[h]
\centering
\begin{subfigure}{.5\textwidth}
  \centering
  \includegraphics[width=83mm]{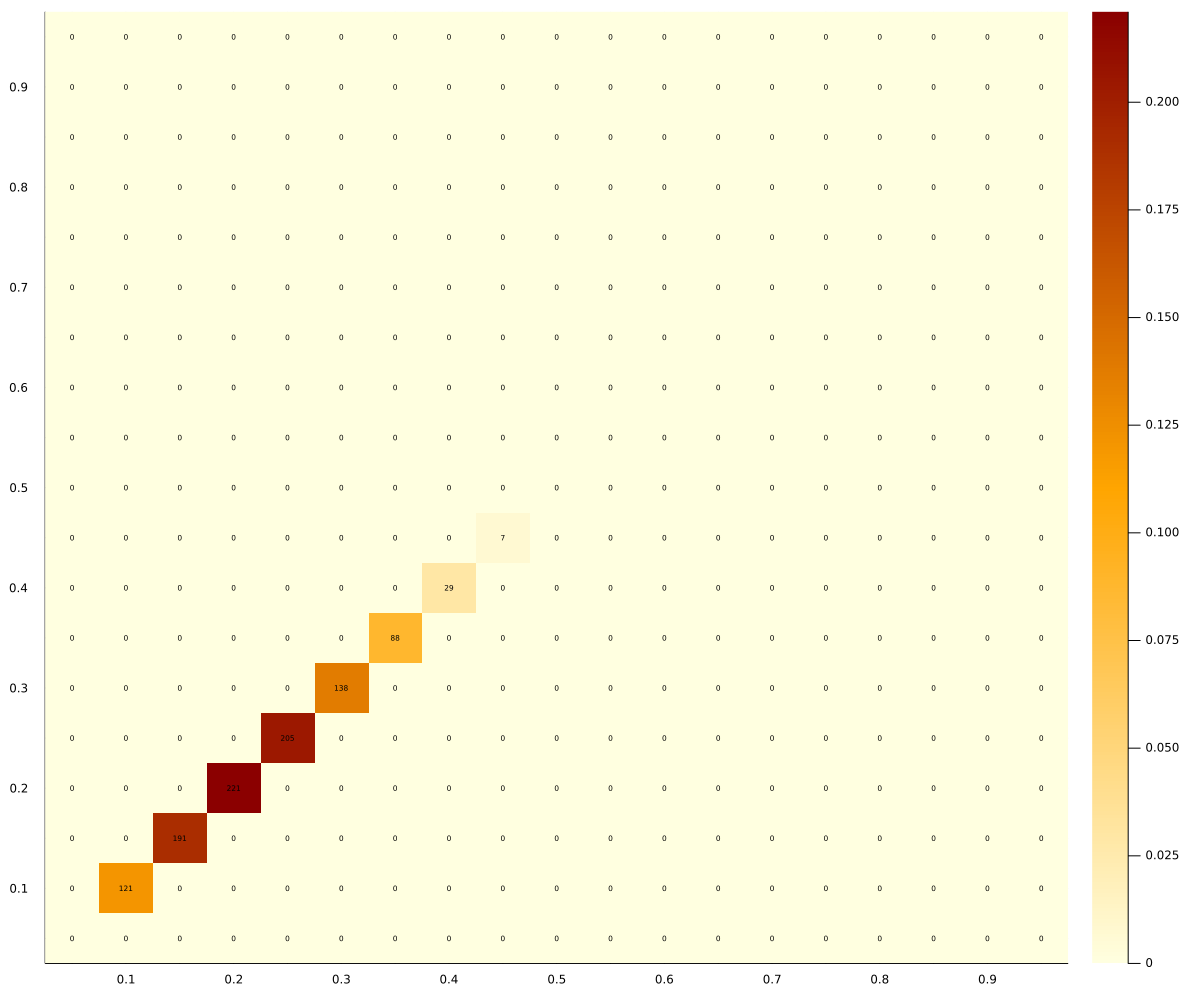}
  \caption{FPA bids}
\end{subfigure}%
\begin{subfigure}{.5\textwidth}
  \centering
  \includegraphics[width=83mm]{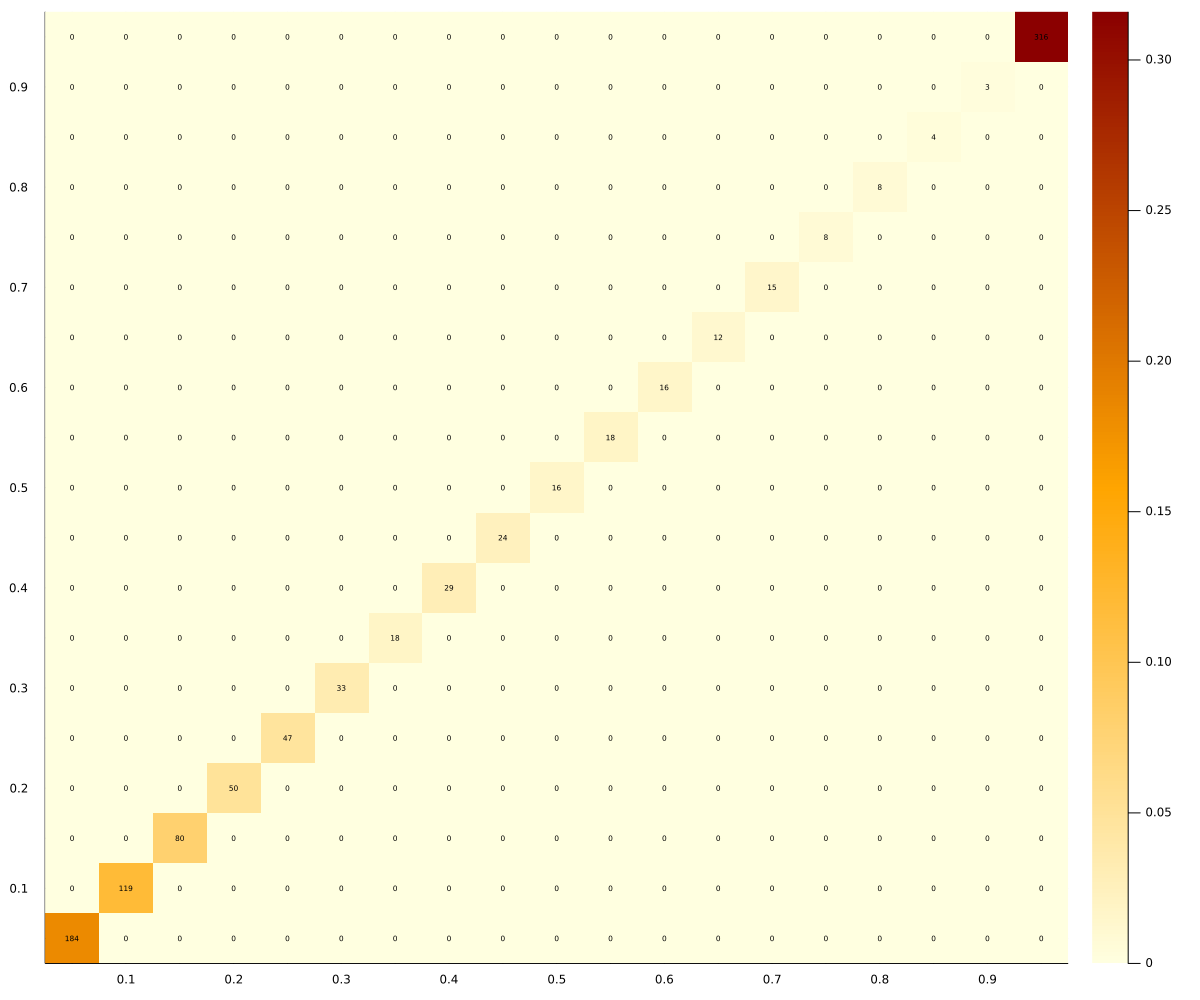}
  \caption{SPA bids}
\end{subfigure}
\caption{Frequencies of bids with a downward force in the exploration rule. $\chi = 0.62 e^{-0.002 t}$.}
\label{fig:pushdown}
\end{figure}

To test this intuition, we introduce an artificial push towards lower bids in the exploration rule. Once the agents reach the steady state, with probability $1-\chi$ they behave according to the $\epsilon$-greedy policy. However, with probability $\chi$ they will choose the lowest bid whose value is ``close'' to the value of the current strategy.\footnote{More precisely, they choose the lowest bid such that the value of that strategy lies within $d = 0.3$ of the value of the current optimal strategy. This translates to an artificial force towards lower bids.} The results of this simulation are summarized in Figure \ref{fig:pushdown}.

\begin{result}
Algorithms designed to explore low bids tend to converge to below-static Nash equilibrium bids in both first and second-price auctions.
\end{result}

As is clear in this Figure \ref{fig:pushdown}, there are still differences between the FPA and SPA, but that is to be expected since there is no way to replicate the same exploration dynamics in the two games. Yet, these results are highly suggestive that this tendency towards local deviations and hence re-coordination on near-by bids is an important force for the difference in outcomes between FPA and SPA. It also explains why in the $\alpha$-price auctions collusive prices slowly emerge as $\alpha$ decreases.

\subsection{How Collusive Prices are Supported}

To understand our results, it is natural to ask how are the low prices maintained in long-term play. In fact, the algorithms do not explicitly try to learn dynamic strategies, such as the ones from our analysis of repeated games, mainly because our algorithms do not keep track of the history of the game --- other than via the $Q$ matrix, which is a very coarse way for keeping history. Moreover, our algorithms are designed to take the best action based on the estimates of long-term payoff consequences and hence they are not designed to execute punishment strategies in case their opponent deviates. 

Admittedly, describing and understanding precisely the evolution of two Q-learning algorithms with 19 different actions each is hard. Based on the analysis of the different simulations and the theoretical work of \citet{Banchio2022b} we have built the following intuition.

Suppose current estimates of $Q$ are such that both bidders choose to bid the same low amount $b$. If there were many periods without exploration, bidders' estimates of $Q(b)$ would converge towards $Q(b)=\frac{1-b}{2(1-\gamma)}$. As a bidder $i$ explores to other bids, it is going eventually observe that a bit higher bid, $b'$, yields on average higher payoffs (since it doubles the probability of winning at a slight increase in payment). That is, eventually $Q_i(b')$ overtakes $Q_i(b)$ and bidder $i$ switches to $b'$ in most periods. That is not stable: the opponent $j'$ estimate of $Q_j(b)$ starts decreasing as a result. At some point it becomes sufficiently low that $j$ starts switching to other bids too. If they switch to $b''<b'$ they will continue losing and reducing $Q_j(b'').$ If they switch to $b'$ they start winning on average half of the time and it may become a new stable profile of bids for a while. If they switch to bidding $b''>b'$ then the instability we reset and now bidder $i$ will after a while start searching for a more profitable bid. 

The key is what happens if they luckily re-coordinate temporarily at $b'$. Now, for bidder $i$ the estimate of $Q_i(b')$ will start decreasing (because they win as often as before at bid $b$, just pay a bit more). The algorithm is agnostic over why that is happening and hence will try to go back to $b$ that had a higher long-term payoff. The same happens to player $j$, hence at some point they will both bid $b$. Additionally, if player $j$ manages to experiment to a lower bid and wins, the players may go back to both bidding $b$ even quicker. If exploration never dies, this cycle can continue forever and include multiple bid levels (it has to include multiple ones).

This is the behavior we observed in our simulation with $100,000,000$ periods and exploration that never stopped (described in Figure \ref{fig:exp}). 

At the essence, two forces affect the FPA equilibrium: one directs bids upwards and the other steers them downwards. In a SPA the latter disappears, leaving only an upward force which naturally leads to convergence on the highest possible pair. 

A slightly different way to see the intuition is that when the players get to the point of both believing that $Q(0.95)$ is the highest and one starts experimenting to lower bids, in SPA there is no reason for the other bidder to ``follow.'' In contrast, in FPA there is a good reason to go down - not to match, but to be just one bid increment above. This makes it easier in FPA than in SPA for the players to escape $(0.95,0.95)$ (or any other high bids) and maintain those lower bids.

This intuition helps understand two more results. First, in the next section we discuss what happens in FPA if the auctioneer gives at the end of each auction additional information about the bids of others. Second, in the section with extensions it helps us resolve why the tacit collusion results in revenues that are meaningfully higher than $0.05$ or $0.10$.

\section{Auction Design.}\label{sec:design}
The results presented so far assume that the designer provides the bare minimum information to the bidders: after each auction she reports solely whether they won or not. However, the designer has access to a wider set of messages: for example, she might want to communicate what the highest bid was, or an anonymized distribution of bids. In this section we show how these design considerations may have a profound impact on the outcomes of play by the bidding algorithms. 

We focus here on first-price auctions to see if the ''collusion'' can be reduced via such a policy. Such auction design policy also has clear practical relevance.

One of the consequences that we now consider is that when an algorithm is provided with information about the highest opponent bid after the auction, it can estimate its payoff in hindsight. That is, it can calculate the counterfactual payoff for every possible bid, not only the one chosen in the auction. This information allows the Q-learning algorithm to learn about all bids contemporaneously. 

\begin{figure}
    \centering
    \begin{subfigure}{.5\textwidth}
        \centering
        \includegraphics[width=83mm]{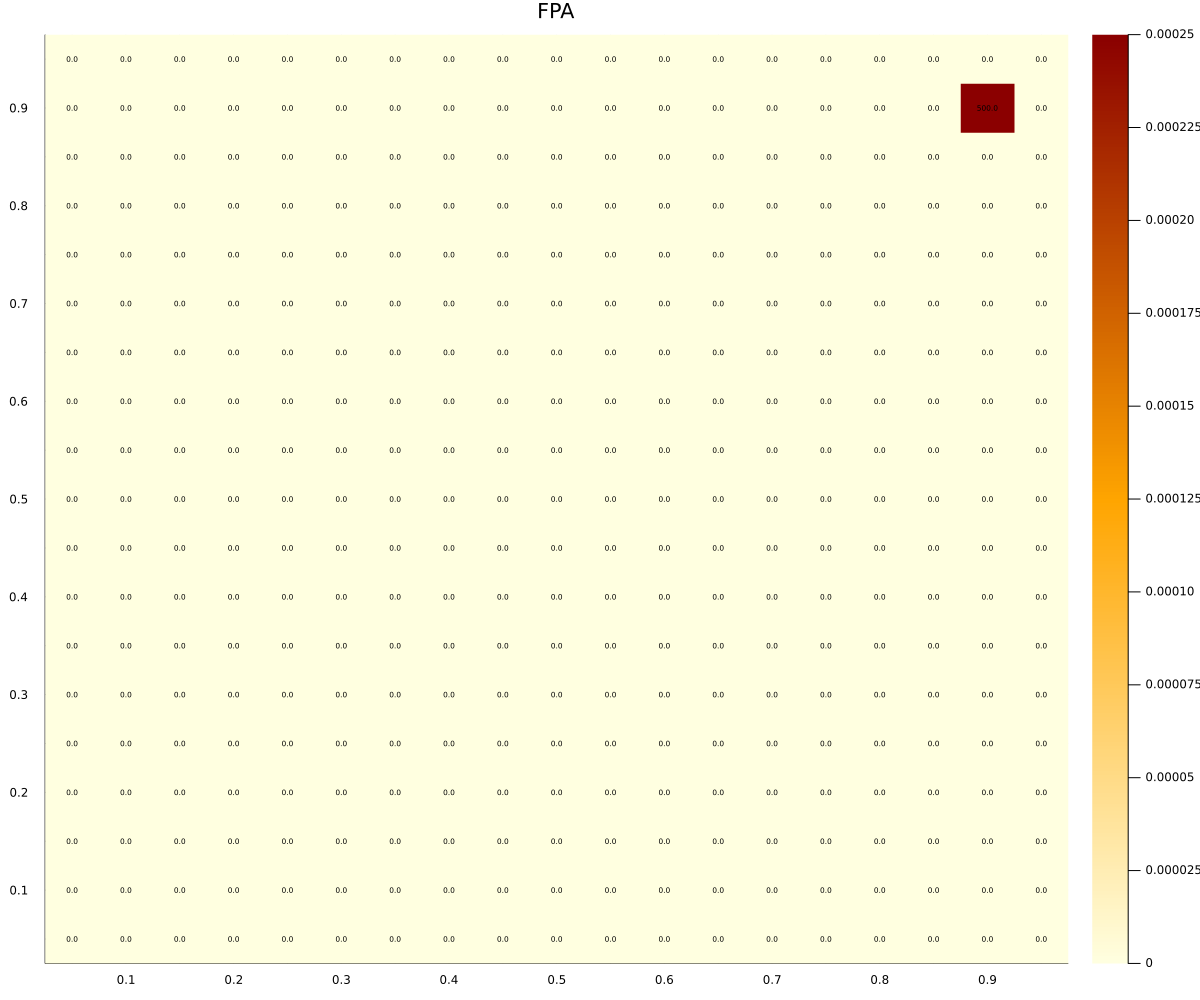}
        \caption{Frequencies of bids}
    \end{subfigure}%
    \begin{subfigure}{.5\textwidth}
        \centering
        \includegraphics[width=83mm]{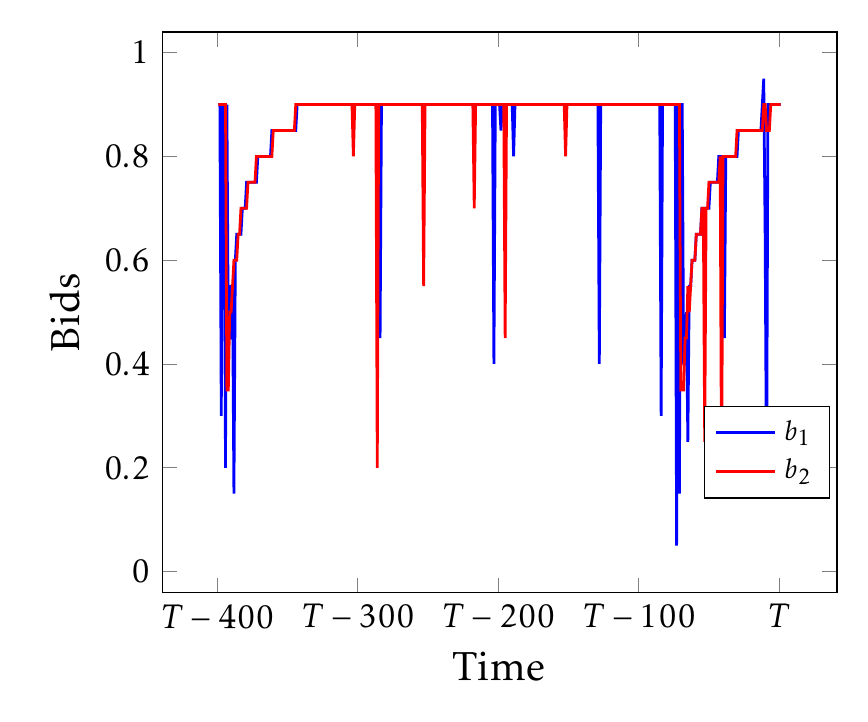}
        \caption{Cycling behavior}
        \label{subfig:cycles}
    \end{subfigure}
    \caption{Outcomes of 500 simulations obtained with the standard set of parameters but synchronous updating.}
    \label{fig:synchronous}
\end{figure}

Our simulations show that with such change in information policy, if our algorithms take that information into account, collusion disappears. In Figure \ref{fig:synchronous} we present the result of simulating a FPA with \emph{synchronous} updating Q-learners. Formally, the algorithms update the action-value function for all entries, using the return in hindsight $R_t(a)$ for each action:
\[Q_{t+1}(a) = (1-\alpha)Q_t(a) + \alpha\big[R_t(a) + \gamma \max_{a'} Q_t(a')\big] \qquad \forall a\]

\begin{result}
The synchronous algorithms (that update $Q$ in hindsight based on the information about the highest competitor bid) in FPA result in substantially higher revenue (close to the static Nash equilibrium) than the asynchronous algorithms described in \ref{Result1}.
\end{result}

The difference in outcomes in Figures \ref{fig:base} and \ref{fig:synchronous} is clear: the algorithms are now competing almost perfectly.\footnote{Competition remains stuck in $(0.9,0.9)$ because of the discretization of the bid space. Finer discretizations approximate competition with higher and higher precision.} This result is consistent with the findings of \citet{Asker2022} in a Bertrand pricing game, where synchronous algorithms compete prices down to marginal cost while asynchronous algorithms collude on supra-competitive prices. 

We can rationalize these outcomes through the mathematical apparatus of \citet{Banchio2022b}. When applying their framework, the synchronous algorithm always suffers from regret towards higher actions, even though the regret is mitigated by the reduction in profit associated with increasing one's bid. The highest regret bid is the one directly above the winning bid: that will be the next preferred action. This process slowly brings the optimal action to the maximum, as shown in Figure \ref{subfig:cycles} that illustrates bid-paths reminiscent of Edgeworth's cycles.\footnote{Such price/bid cycles have been often documented in algorithmic pricing and auctions (see for example \citet{Musolff2021}, \citet{Edelman2007}).} Experimentation forces the algorithms to try a dominated action, and sometimes this leads to both bidders coordinating on lower bids. However, the regret process described before soon brings both bidders back to the competitive state.

A related intuition is related to what we discussed above: For re-coordination on low bids it is important that when bidders switch to $b'>b$ and then coordinate on $b'$, over time they understand that $b$ was actually better. But with synchronous learning they update their estimate of the value of $b$ conditional on their opponent playing $b'$: their value for $b$ is bound to fall. The only hope for recoordination are second-order events, when both players experiment at the same time. 

The analysis just carried out turns out to be an important design consideration in online advertising. Recently, Google changed their ad auctioning system from a SPA to a FPA. Alongside this change, bidders now observe the highest bid of their competitor. This fact squares perfectly with the intuition we obtained from our simple simulations: the ability to compute regret introduces an incentive to outbid the opponent that is missing otherwise. This is strong evidence that design choices such as communication and information may have a large impact on outcomes and consequently on revenues.\footnote{See \citet{dworczak2020mechanism} for other reasons why information disclosure after an auction can affect revenues and efficiency.}

\section{Extensions.}
\begin{figure}[h!]
\centering
\begin{subfigure}{.5\textwidth}
  \centering
  \includegraphics[width=83mm]{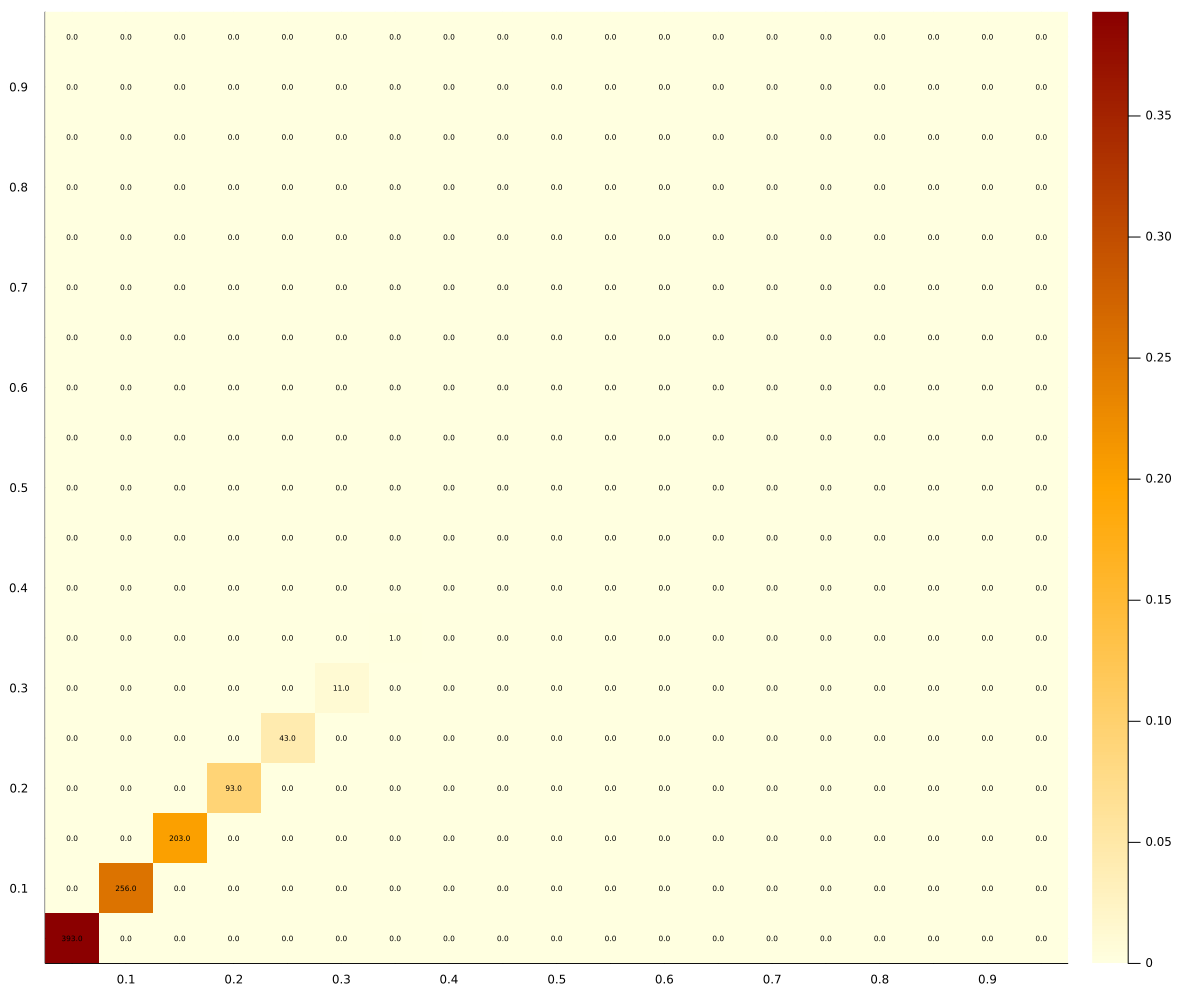}
  \caption{FPA bids}
\end{subfigure}%
\begin{subfigure}{.5\textwidth}
  \centering
  \includegraphics[width=83mm]{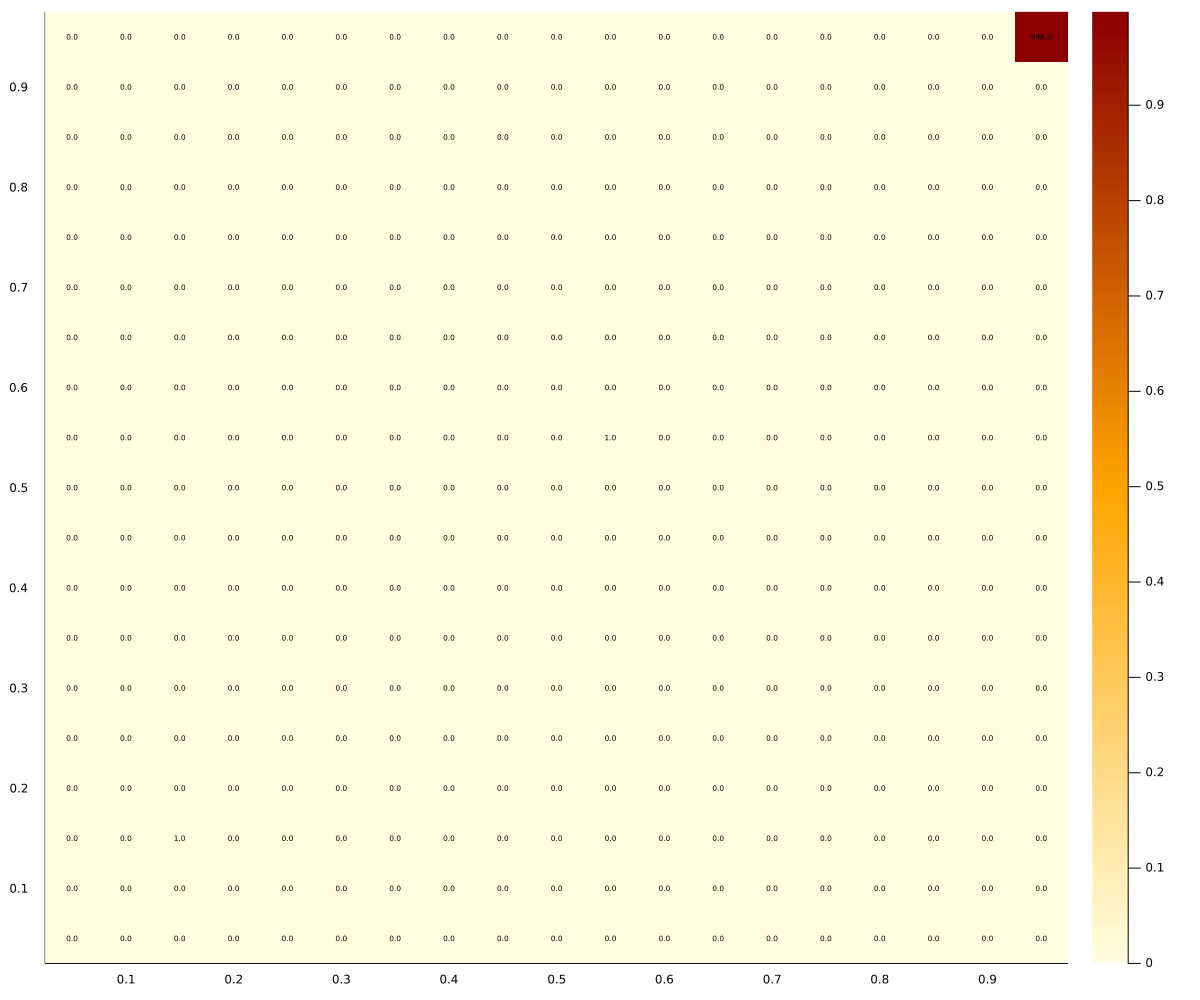}
  \caption{SPA bids}
\end{subfigure}
\caption{Frequencies of bids excluding all the additional negative bids (simulations never converge to the negative bids).}
\label{fig:nobid}
\end{figure}

One puzzle remains open, when observing the results in Figure \ref{fig:baseFPA}. While collusion is effective at improving the bidders' payoffs, it is imperfect: why are the bidders colluding on average on $(0.3,0.3)$ instead of maximizing their profit by coordinating on $(0.05,0.05)$? The explanation turns out to be rather mechanical. In fact, if the algorithms were choosing a price at random, then a bid of $0.5$ would be profitable $50\%$ of the time, while a bid of $0.05$ only $5\%$. In order to collude on bids of $0.05$, the two algorithms would need to each bid $0.05$ repeatedly and split the surplus, which is rather unlikely. To test this theory, we run an experiment with additional negative bids, between $-0.3$ and $0$. These bids can be thought of as non-participation options. None will clearly ever be chosen in equilibrium, but now, when bidders are choosing at random, a bid of $0.05$ is profitable almost $30\%$ of the time. The results of this experiment are shown in Figure \ref{fig:nobid}

\begin{result}
The algorithms find it easier to coordinate on the lowest bid in a first-price auction when they are given the option to not participate.
\end{result}

As it turns out, collusion now occurs more frequently on the most profitable pairs, validating our earlier explanation.

\begin{figure}[h]
\centering
\begin{subfigure}{.5\textwidth}
  \centering
  \includegraphics[width=83mm]{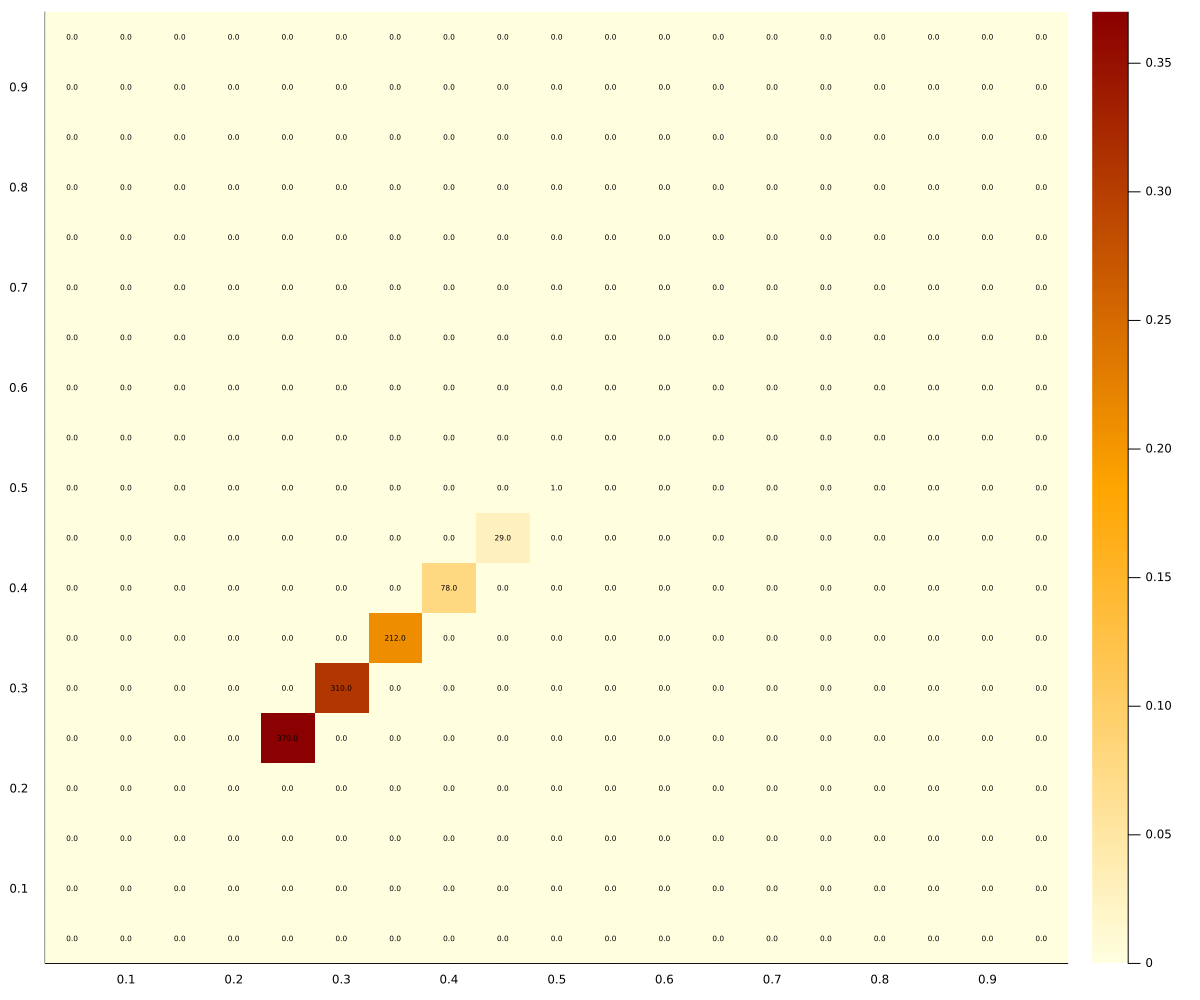}
  \caption{FPA bids}
\end{subfigure}%
\begin{subfigure}{.5\textwidth}
  \centering
  \includegraphics[width=83mm]{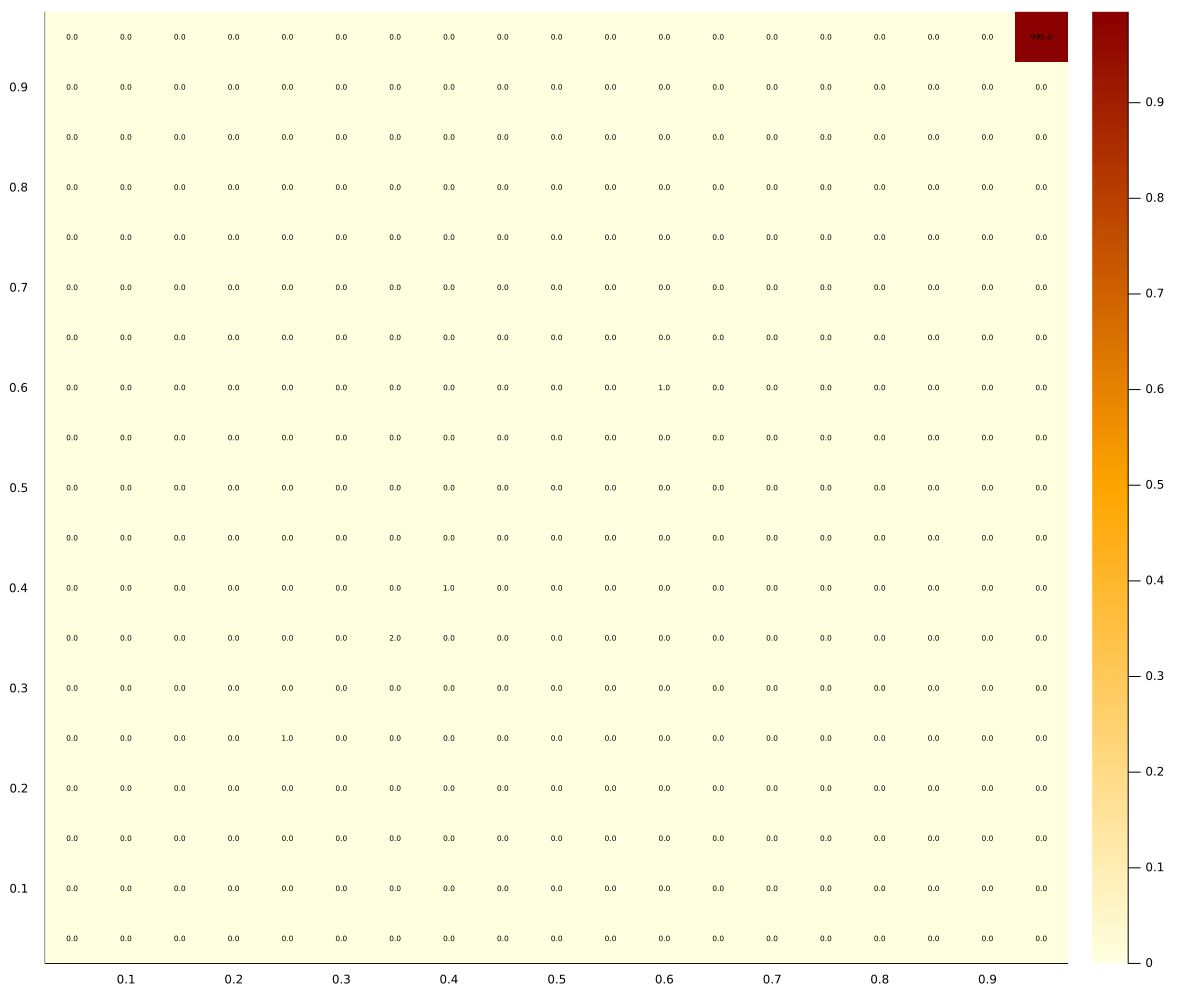}
  \caption{SPA bids}
\end{subfigure}
\caption{Frequencies of bids with a reserve price of $0.2$.}
\label{fig:reserve}
\end{figure}

\paragraph{Reserve prices.} 
From the standard theory of auctions we expect reserve prices to matter only when they are binding. In principle, then, the SPA should be unaffected and the FPA should be essentially unaffected as well as long as the reserve is lower than the collusion pairs. To test this, we run an experiment with a reserve price $r = 0.2$. The results are shown in Figure \ref{fig:reserve}

\begin{result}
With reserve prices, the algorithms coordinate on a distribution that appears skewed towards higher bids.
\end{result}

These graphs conform well with our intuition: the two distribution of outcomes in the FPA with reserve is close to a truncation of the original  distribution. Note that this also relates to our previous observation: the bids below the reserve price act as ``non-participation'' bids, helping bidders coordinate on the most-profitable outcome. We will show that the introduction of a fringe of bidders instead has additional effects on the outcome distribution.

\paragraph{More competition.} Our model so far has been dealing with 2 bidders competing against each other. The results hold with 3 bidders as well, as shown in Figure \ref{fig:three}.
\begin{result}
With three bidders, in a first-price auction the algorithms are more likely to converge on collusive outcomes for higher values of the discount factor. In the second-price auction the outcome remains the static Nash equilibrium. 
\end{result}
However, note that for the discount factor used so far ($\gamma = 0.99$) collusion is harder to sustain with three bidders. If we run the experiment with $\gamma = 0.999$ however, collusion is restored.

\begin{figure}[h]
\centering
\begin{subfigure}{.5\textwidth}
  \centering
  \includegraphics[width=83mm]{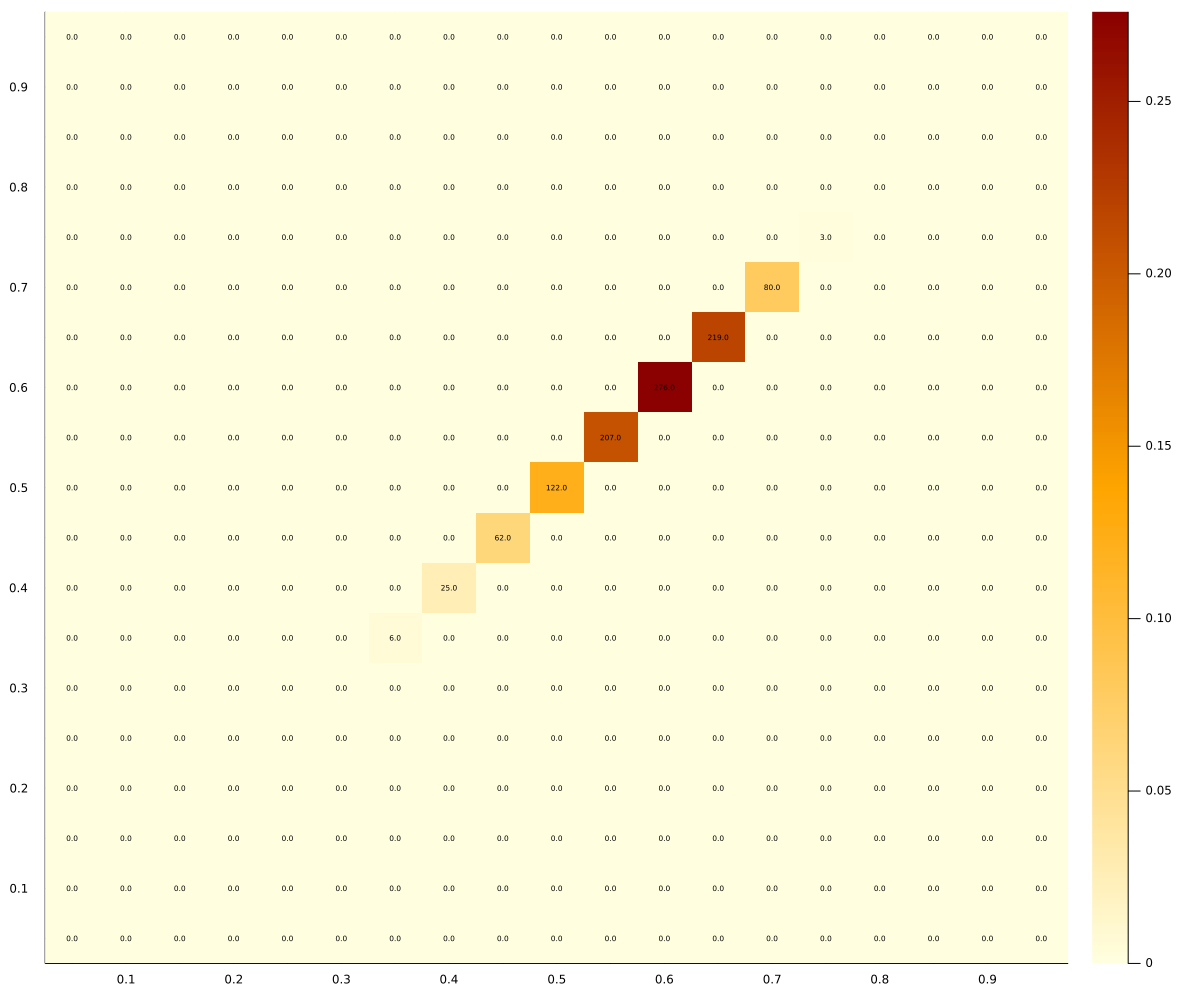}
  \caption{FPA bids}
\end{subfigure}%
\begin{subfigure}{.5\textwidth}
  \centering
  \includegraphics[width=83mm]{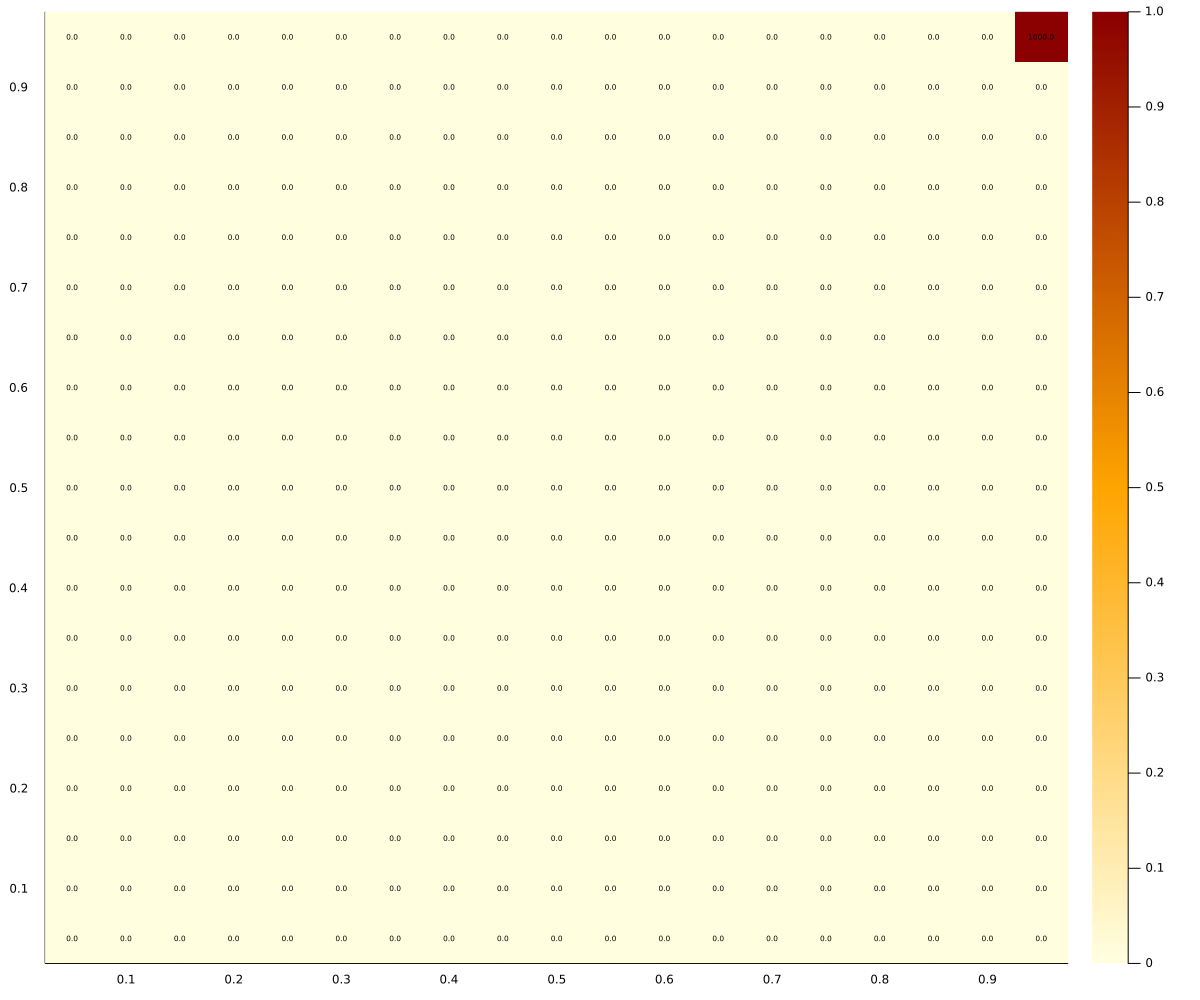}
  \caption{SPA bids}
\end{subfigure}
\caption{Frequencies of bids with a Uniform$(0,1)$ fringe.}
\label{fig:fringe}
\end{figure}

One further way to add competition is to add a fringe of non-strategic bidders. We model the fringe as an additional bid randomly drawn from a uniform distribution over the unit interval. The outcomes of the experiments are presented in Figure \ref{fig:fringe}.

\begin{result}
With a fringe of bidders with bids drawn from a uniform $[0,1]$ distribution, the bids in the first-price auction increase over the optimal best-response. In the second-price auction the outcome remains the static Nash equilibrium.
\end{result}

In particular, notice how in the FPA the bids go up, mostly above the optimal collusion above the fringe. Given the values of the experiment, the optimal bid for both agents would be $0.5$. Instead, we observe a distribution that concentrates the most mass on $0.6$ and $0.65$, thus reducing the gains from collusion significantly.

\section{Conclusions.}
We have analyzed simple auctions where bidders with fixed values compete repeatedly in many auctions. Instead of using Nash equilibrium analysis, we asked how play would evolve if two (and sometimes more) bidders used simple artificial-intelligence algorithms. We simulated simple Q-learning algorithms and have shown a remarkable difference in the performance of first-price and second-price auctions.

This difference points to a new intuition about the play of such algorithms in real-life settings. Our previous intuitions are built either on the expectation that simple algorithms could converge to the static Nash equilibrium (and then, our Nash equilibrium analysis would predict no significant difference in the revenues), or on the expectation that these algorithms will coordinate on a tacit-collusive equilibrium of the repeated game. In the latter case, the theory of repeated games again would not predict differences between the first and second-price auctions. As we discussed, the incentive compatibility constraints for collusive strongly symmetric equilibria are approximately the same in those two formats (they are different only because we assume bids have to be chosen from a grid). Moreover, when one considers asymmetric equilibria (like bid rotation, where bidders alternate who wins), tacit collusion is much easier to sustain in repeated second-price auctions. 

We show that a new dynamic force creates the observed difference in outcomes: in a second-price auction any deviation to a bid higher than the competitor’s bid is equally profitable. In the first-price auction, the most profitable deviation is by just one bid increment more than the opponent. As a result, when because of experimentation bidders deviate from equally low bids, the losing algorithm starts exploration that can only end (at least temporarily) when both bidders re-coordinate and believe that winning half of the time at this bid is better (for long-term payoffs) than deviating to a higher bid.
Such re-coordination happens at lower bids in the first-price auction than in the second-price auction, resulting in different behavior in the long run. 

We note that the seeming convergence to low bids is qualitatively different than the convergence to the static Nash equilibrium. For any fixed action of an opponent, our algorithms learn to best respond. Hence, if they converge to a fixed profile of actions then it must be a static Nash equilibrium. This is what happens in the SPA and in the FPA when the auctioneer provides them information that facilitates synchronous learning --- estimating the value of all bids in every round. 
However, without that information, in the first-price auction the bidders never properly converge: intuitively they end up in local cycles: for example, they get to bidding 0.3 each, then one of them experiments and they learn that a higher bid is more profitable - they may learn that 0.35 is a best response - and then a short phase of experimentation takes place, where the opponent tries different actions to counteract the streak of losses. After re-coordinating on (0.35, 0.35), over time they learn that their average payoff is worse than the payoff from 0.3, and they try to get back to 0.3. With some luck, they both try lower bids at the same time and they learn that indeed that is a better strategy and switch back to 0.3 (or to even lower bids). In other words, especially when we look at the system without taking the experimentation parameter $\epsilon$ to zero, the two players spend most of the time on the  diagonal (with 
equal bids), but do not settle in one place forever. Instead they move up and down. 

If the auctioneer provides them with information about the lowest bid to win, when they move from the bid of 0.3 to 0.35, they synchronously learn that while 0.3 used to give a better payoff than 0.35, given the current state of the system, 0.35 is actually better than 0.3, and the incentive to go back to 0.3 disappears. That explains why providing this additional information (if it is not ignored by the algorithms or used in some other way than in our simulations) leads to more competitive bidding.

Many questions remain open. First, one may be worried about the robustness of our findings to allowing other artificial intelligence algorithms to play these games. We expect that the new force we have identified will be present in many algorithms that operate with limited information (for example, in first-price auctions, without observing bids of others). 
Related to that question of other algorithms is what would happen if we made the Q-learning algorithms more sophisticated, for example, by keeping as a state whether the bidder won or lost the last auction (or what fraction of auctions they have won in the last minute). 

Second, we have considered the policy of revealing additional information in first-price auctions by allowing the algorithms to update the Q vector synchronously but otherwise we kept the algorithm unchanged. A realistic concern rooted in economic theory of repeated games is that providing additional information could facilitate tacit collusion - bidders could switch to algorithms that keep last two bids in short-term memory as states and estimate a Q matrix, with each vector representing a different pair of recent bids. If so, providing information could backfire by facilitating instead of discouraging tacit collusion (for example, in the form of bid rotation). 

Third, we looked at a very simple environment with two symmetric bidders and fixed values. While we have shown robustness of our findings to the introduction of a third bidder and a competitive fringe, many questions remain open. We think the most interesting question is how asymmetry would affect the findings (asymmetry in values and/or algorithms). Time-varying valuations are also of great interest since they would provide additional rationale for using artificial intelligence algorithms that constantly experiment and try to adapt to the changing competitive environment.

\bibliographystyle{ecta} 
\bibliography{bibliography}

\begin{appendices}

\section{Repeated Games Equilibria}\label{app:equilibria}
In this appendix we discuss the collusive Nash equilibria of the game of Section \ref{sec:model}.

\paragraph{Strongly Symmetric Equilibria.} 
One class of equilibria is Strongly Symmetric (Subgame Perfect Nash) equilibria. In such equilibria, bids of the players are symmetric after every history of play. Following standard arguments, in this perfect-monitoring game, the best strongly symmetric equilibrium has the bidders submit bids $b_1$ (the smallest allowed bid) and win with probability $\frac{1}{2}$. These bids continue as long nobody deviates. Upon deviation, players forever switch to the static Nash equilibrium with bids $b_m$.

This pair of strategies forms an equilibrium of a repeated first-price auction if and only if:
\[\frac{1-b_1}{2(1-\gamma)}\geq 1-b_2 +\gamma \frac{1-b_m}{2(1-\gamma)},\]
where the left-hand side is the long-term profit from bidding $b_1$, and the right-hand side is the short-term profit of a marginal increase in one's bid followed by $b_i=b_m$ forever after. The equilibrium condition above can be rearranged as 
\[\gamma \geq \gamma^*_{FPA} = \frac{m-2}{2m-3}. \]
Recall that $m$ is the discretization parameter: when $m$ is large the critical discount factor converges to $\gamma \geq \frac{1}{2}$.

In a second-price auction, the inequality is somewhat harder to satisfy, because a one-shot deviation is slightly more profitable (due to the grid on available bids). The analog incentive-compatibility is: 
\[\frac{1-b_1}{2(1-\gamma)}\geq 1-b_1 +\gamma \frac{1-b_m}{2(1-\gamma)}.\]

This reduces to 
\[\gamma \geq \gamma^*_{SPA}=\frac{m}{2m-1}.\]
As $m$ grows large, this critical threshold also converges to $\frac{1}{2}$. 

In summary, for any discretization $m$, the threshold is always lower for a first-price auction: $\gamma^*_{FPA}<\gamma^*_{SPA}$, but the difference is negligible for large $m$. This analysis suggests that it may be easier to collude tacitly in an FPA, but the differences should be minor.\footnote{Following the terminology in the literature on bidding in repeated auctions, by ''tacit collusion'' we mean equilibria with revenues smaller than in the repetition of the static Nash equilibrium.}

\paragraph{Bid Rotation.} 
The strongly symmetric equilibria we described require observing both bids. With asymmetric equilibria, tacit collusion may be even easier to sustain and require even less information for monitoring.

A bid rotation scheme (BRS) works as follows: bidders take turns between winning and losing each auction. For example, bidder 1 is supposed to win all auctions in odd periods, and bidder 2 in even periods.\footnote{See \citet{Mcafee1992} and \citet{Skrzypacz2004}  for further discussion of bid rotation schemes.} The bidder that is supposed to lose bids the smallest possible amount, $b_1$. Deviations that lead to the wrong bidder winning are followed by a permanent deviation to bidding $b_m$, the repetition of the static Nash equilibrium.\footnote{One may be skeptical how the players could tacitly coordinate on such an odd-even split without direct communication. A perhaps more realistic equilibrium would have bidders bid symmetrically in the first auction, and afterward, the winner in a previous auction would let their opponent win in the current auction and so on.} 

To be more specific, given our grid of allowable bids, the BRS works as follows in the FPA and SPA. 

In a FPA, bidder $1$ bids $b_2$ (one bid increment above the lowest bid) in odd periods and $b_1$ in even periods (the lowest bid, to lose), while bidder $2$ does the opposite (observable deviations lead to forever reversions to $b_m$).\footnote{The way we wrote the game and ran simulations, we forced the bidders to bid at least $b_1$ in every auction. When the grid is fine, that may not be an important assumption. In one of our simulations, bidders could choose not to bid at all in any given period.}
When considering deviations, the bidders trade off future large discounted profits in every other period with an immediate payoff followed by limited profits forever after.
For this BRS to be an equilibrium of the repeated FPA the following incentive compatibility condition must be satisfied (this is also a sufficient condition): 

\[\gamma\frac{1-b_2}{1-\gamma^2} \geq 1-b_2 +\gamma \frac{1-b_m}{2(1-\gamma)}.\]

It simplifies to:

\[\gamma \geq \frac{1}{2}\sqrt
{\frac{10m-11}{2m-3}} - \frac{1}{2}.\]

If we take $m$ to infinity, it converges to $\gamma \geq \frac{\sqrt{5}-1}{2}=0.62$.
Note that in FPA, this condition is more stringent than the condition for the strongly symmetric equilibrium. There are two reasons for it. First, this collusive equilibrium is less profitable for a finite grid than the strongly symmetric equilibrium (the winner pays $b_2$ instead of $b_1$). That difference disappears in the limit as $m$ gets large. 
Second, even in the limit, the incentive compatibility constraints are harder to satisfy in BRS. In BRS, when a bidder is supposed to lose, a deviation increases the probability of winning from $0$ to $1$. In the strongly symmetric equilibrium, a deviation increases the probability of winning only from $\frac{1}{2}$ to $1$.

In a SPA, a BRS can work even better. The strategies in the best (in the sense of easiest-to-satisfy incentive compatibility constraints) are different than in the FPA. Bidder 1 bids $b_1$ in even periods and $b_m$ in odd periods (while bidder $2$ does the opposite). Observable deviations (when the wrong player wins) are punished by reversing to $b_m$ forever (as before). 
The critical difference is that the player expected to win bids the closest to their value. It does not cost the players higher payments in a SPA, but it would in an FPA. Such bidding helps sustain the BRS as an equilibrium in SPA because a deviating player would have to pay the high bid.

The (necessary and sufficient) indifference condition for the BRS in SPA is:
\[\gamma\frac{1-b_1}{1-\gamma^2} \geq \frac{1-b_m}{2} + \gamma\frac{1-b_m}{2(1-\gamma)}.\]
This simplifies to: 
\[\gamma \geq \frac{1}{2m-1},\]
and in the limit, as $m$ gets large, it converges to $\gamma \geq 0.$ 

The intuition for that (perhaps surprising) result is that in the limit with a continuum of bids, one player bidding $v_i$ and the other player bidding $0$ is a Nash equilibrium of the static game. So no dynamic punishments are necessary to sustain BRS in the repeated game.

\section{Additional Figures}

\begin{figure}[h!]
\centering
\begin{subfigure}{.5\textwidth}
  \centering
  \includegraphics[width=83mm]{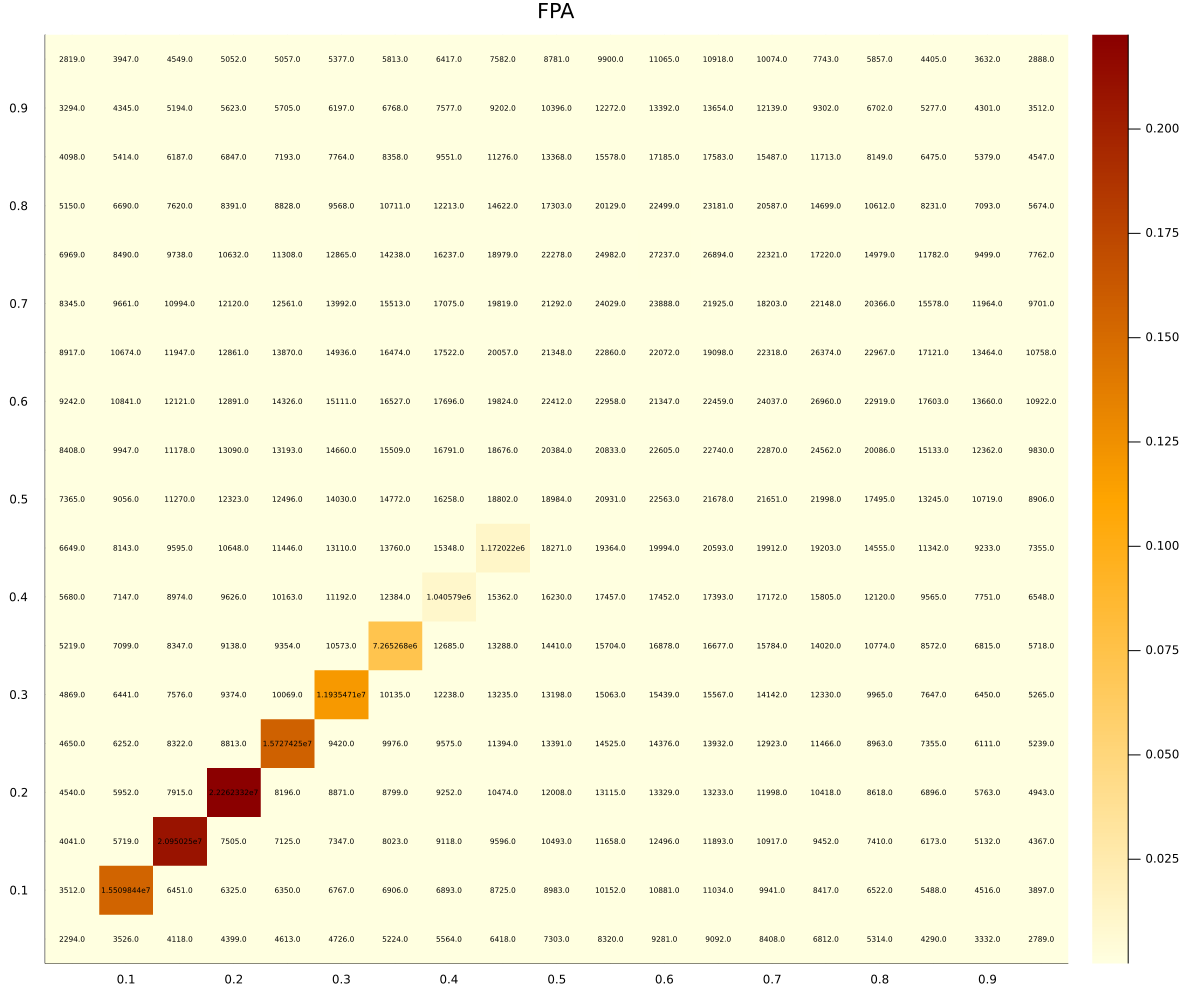}
  \caption{FPA bids}
\end{subfigure}%
\begin{subfigure}{.5\textwidth}
  \centering
  \includegraphics[width=83mm]{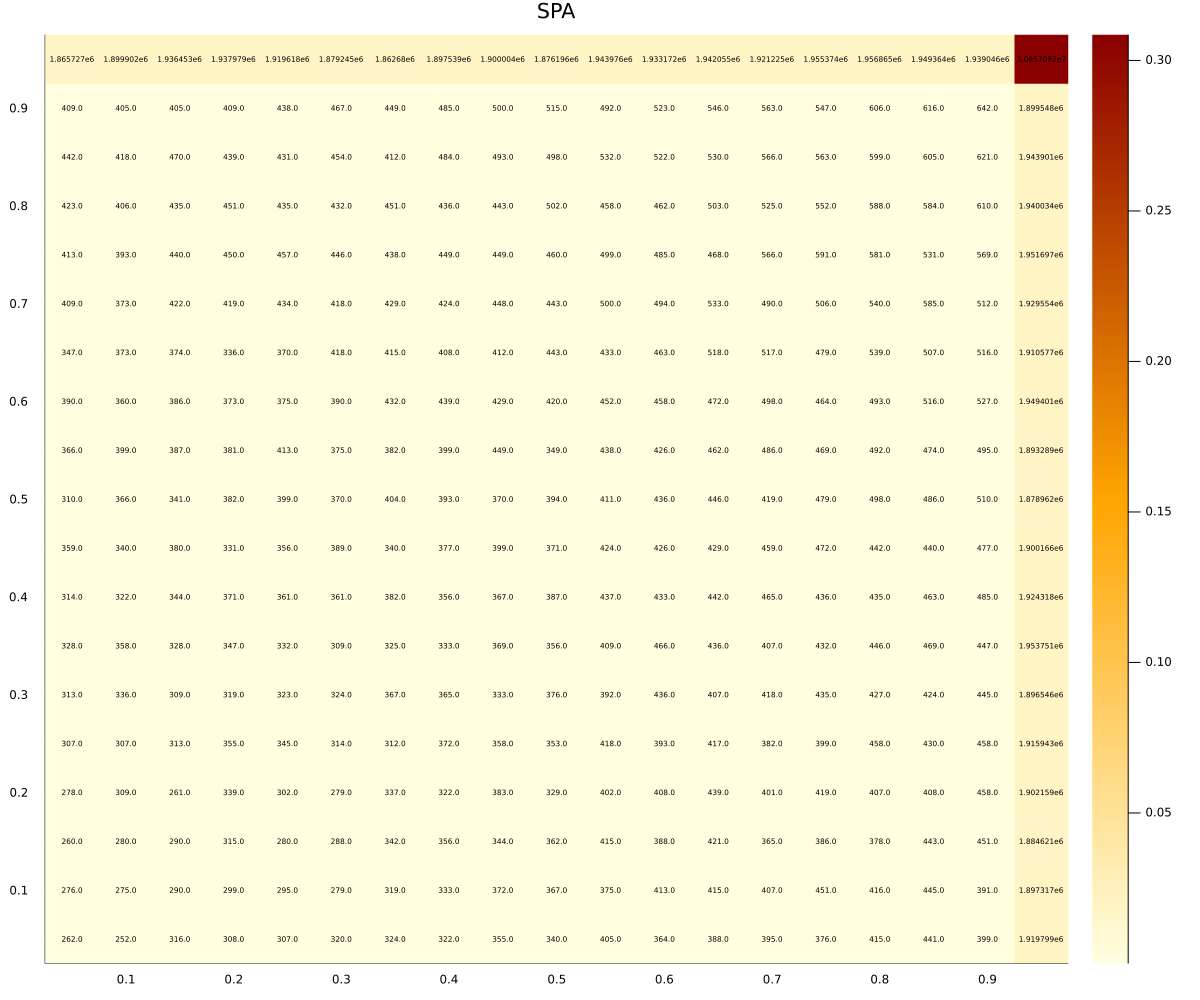}
  \caption{SPA bids}
\end{subfigure}
\caption{Frequencies of bids from one simulation with 100,000,000 iterations and continuous exploration. Parameters: $\varepsilon = 0.001, \beta=0.$}
\label{fig:exp}
\end{figure}

\begin{figure}[h!]
\centering
\begin{subfigure}{.5\textwidth}
  \centering
  \includegraphics[width=83mm]{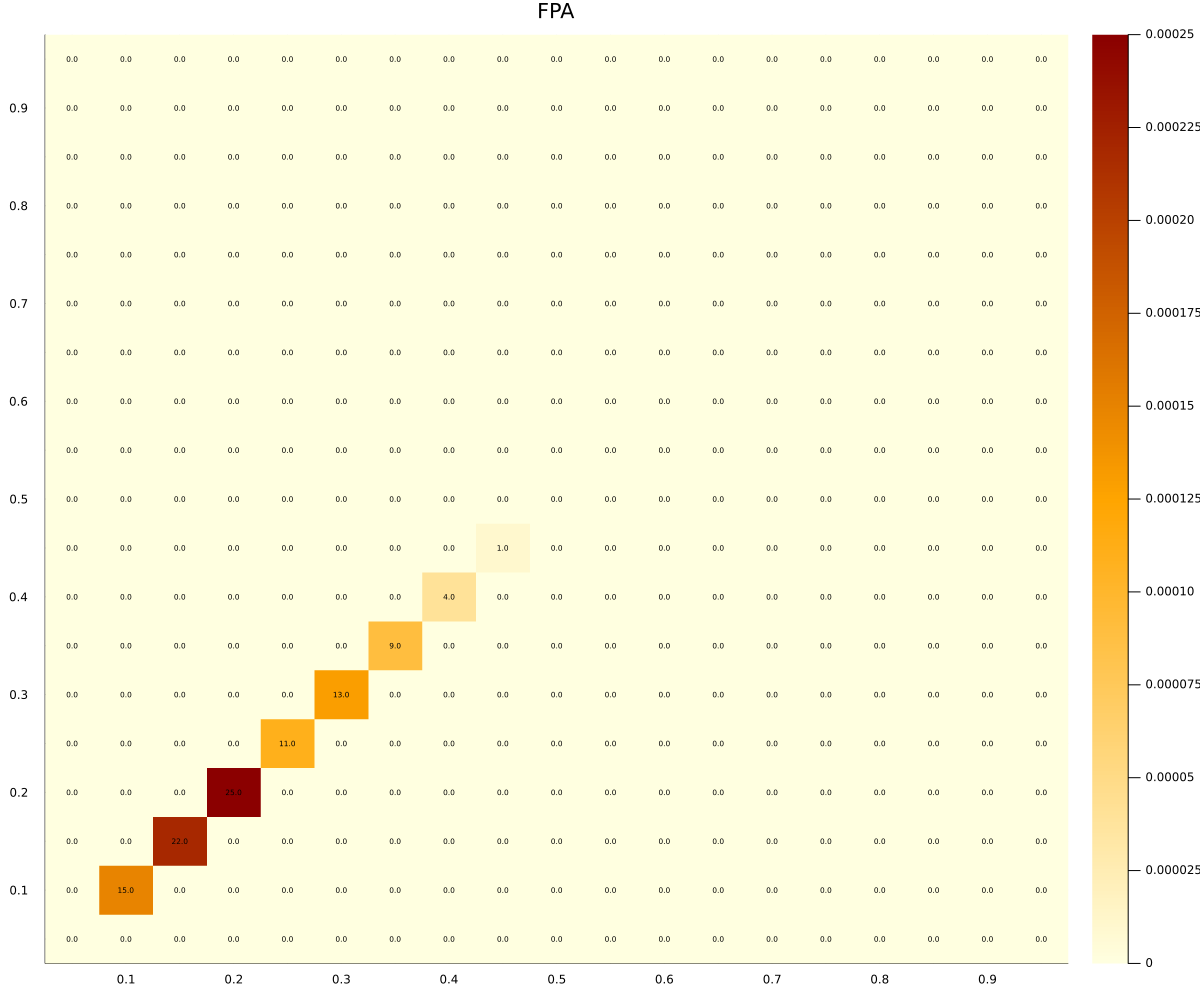}
  \caption{FPA bids}
\end{subfigure}%
\begin{subfigure}{.5\textwidth}
  \centering
  \includegraphics[width=83mm]{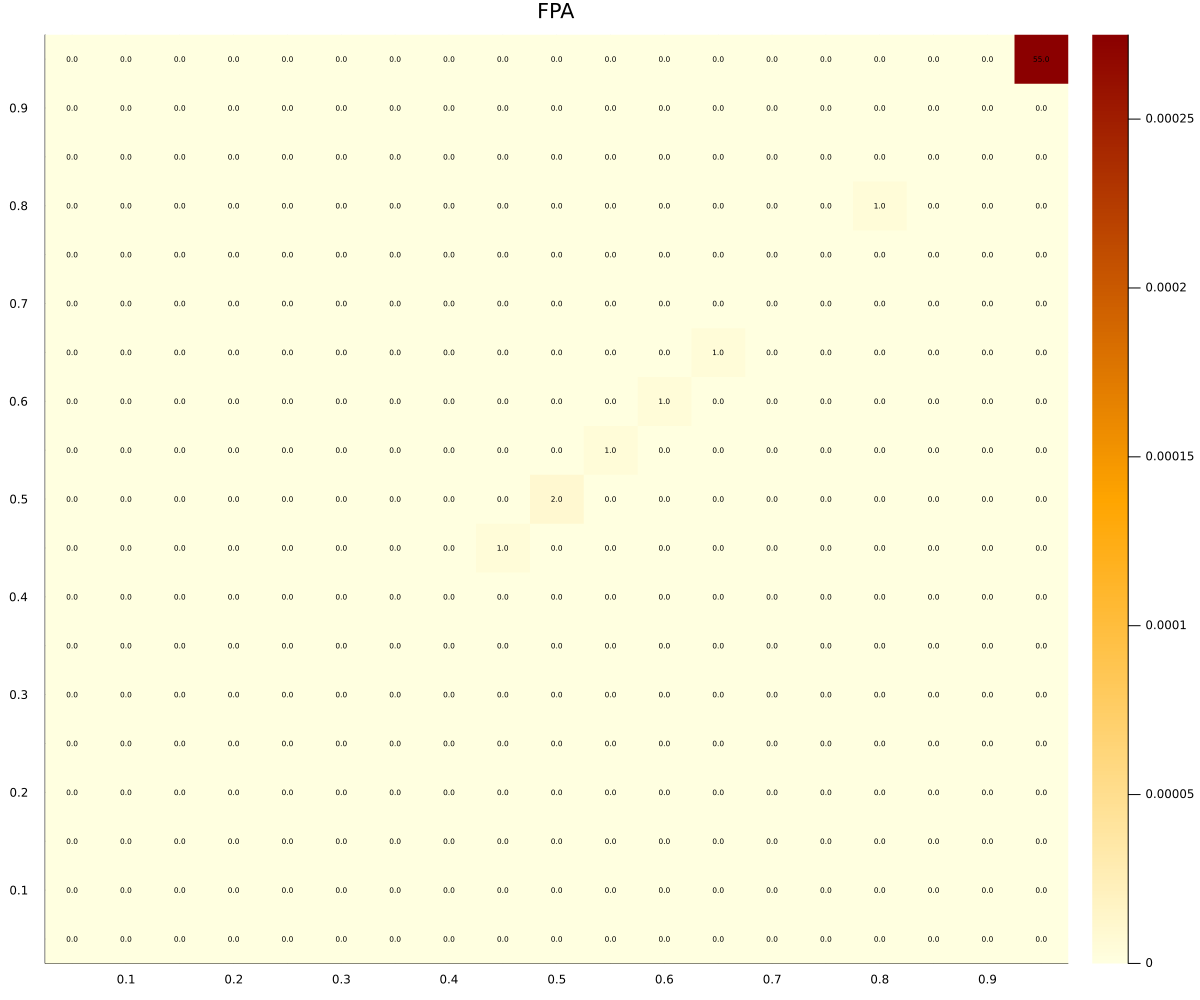}
  \caption{SPA bids}
\end{subfigure}
\caption{Outcomes of 100 simulations with algorithms limited to local exploration.}
\label{fig:local}
\end{figure}

\begin{figure}[h!]
\centering
\begin{subfigure}{.5\textwidth}
  \centering
  \includegraphics[width=83mm]{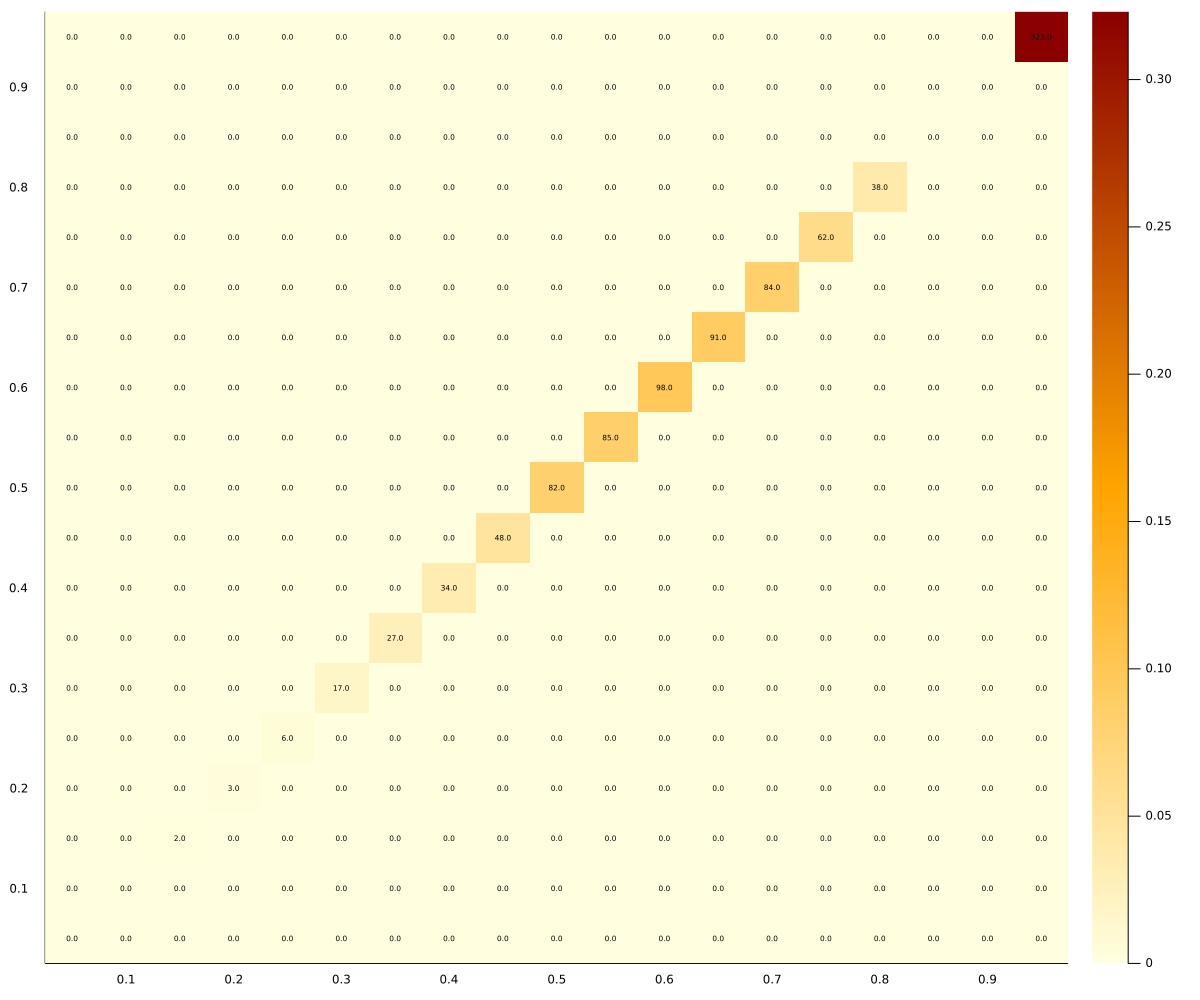}
  \caption{FPA bids, $\gamma = 0.99$}
\end{subfigure}%
\begin{subfigure}{.5\textwidth}
  \centering
  \includegraphics[width=83mm]{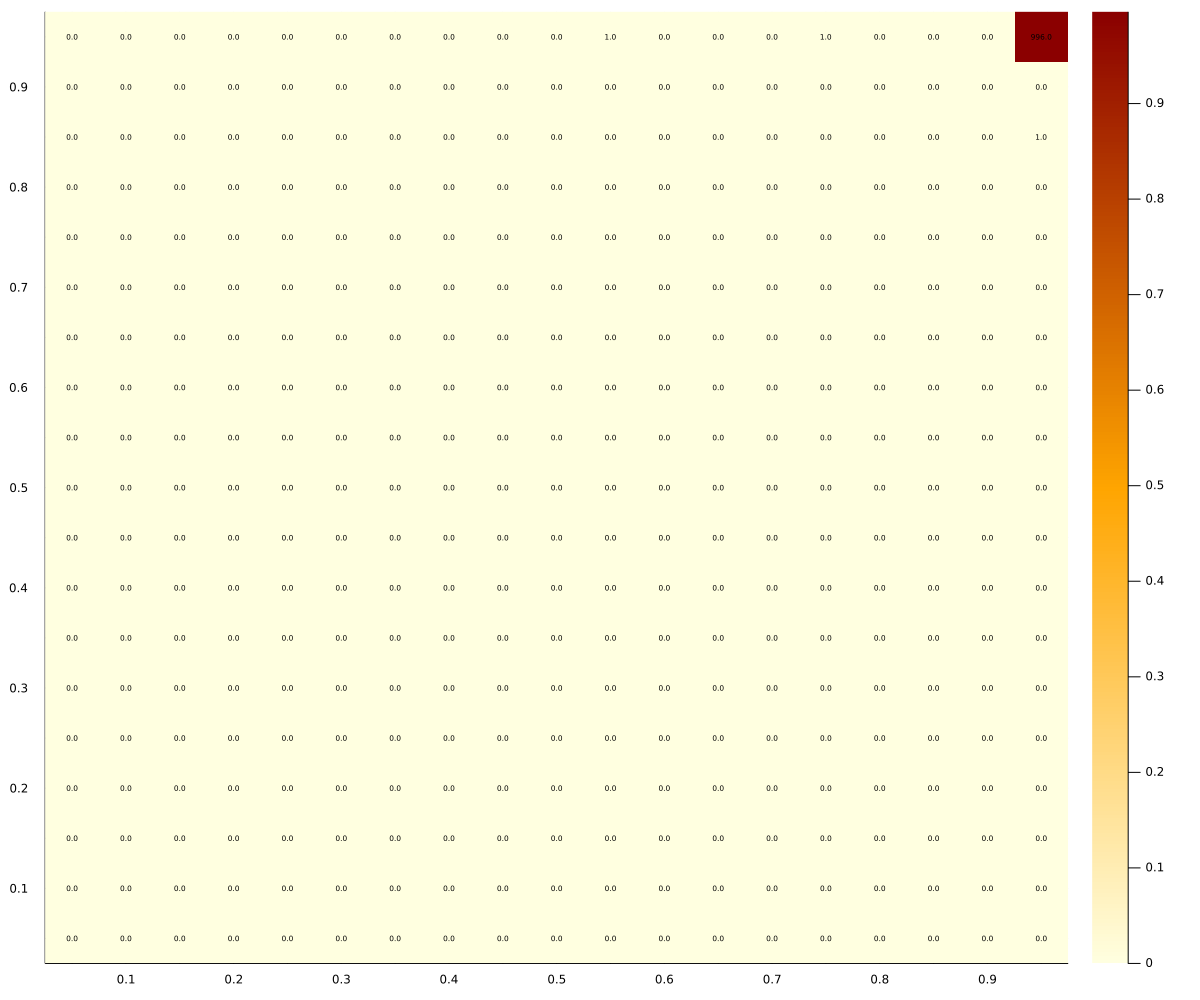}
  \caption{SPA bids, $\gamma = 0.99$}
\end{subfigure}
\begin{subfigure}{.5\textwidth}
  \centering
  \includegraphics[width=83mm]{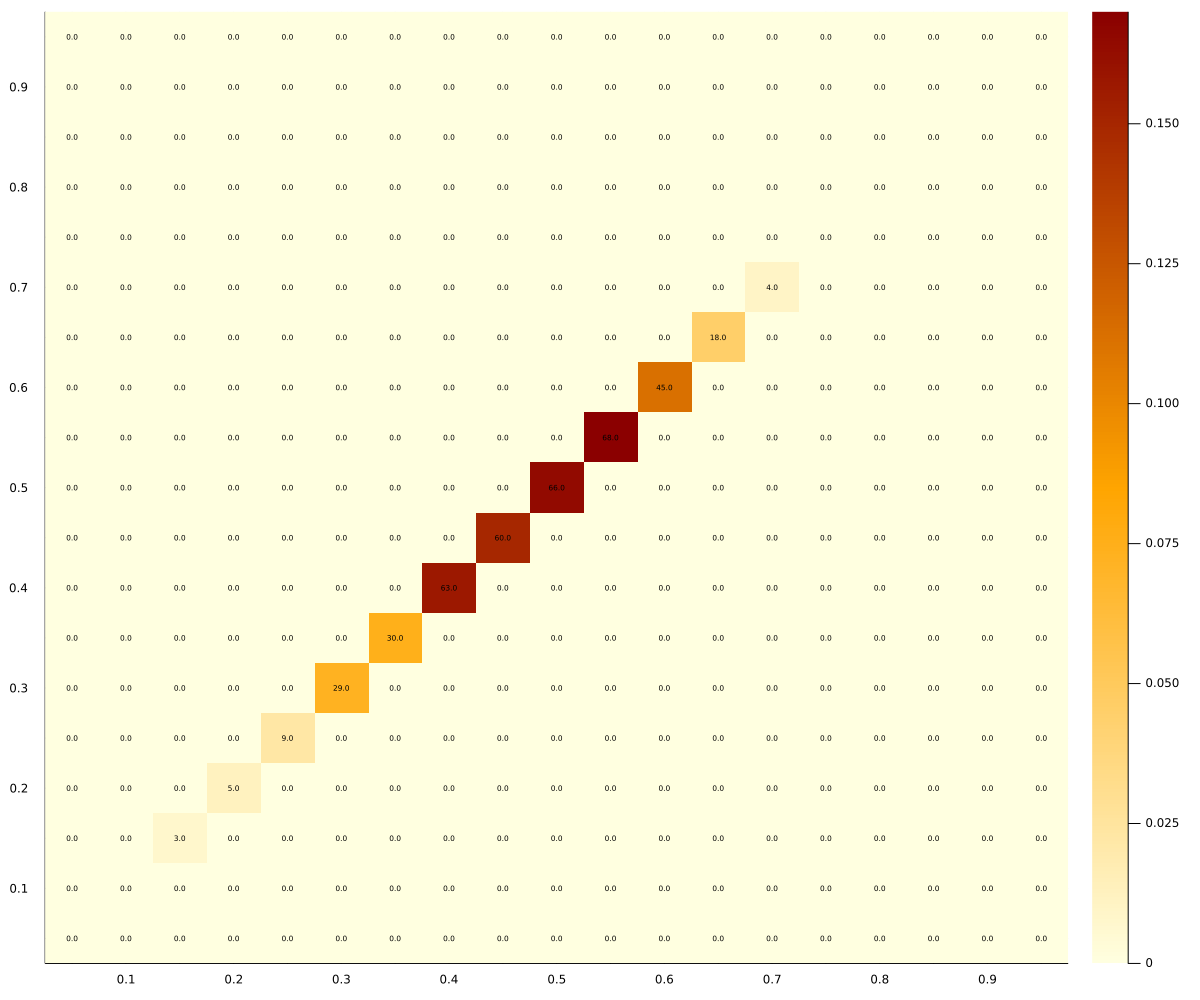}
  \caption{FPA bids, $\gamma = 0.999$}
\end{subfigure}%
\begin{subfigure}{.5\textwidth}
  \centering
  \includegraphics[width=83mm]{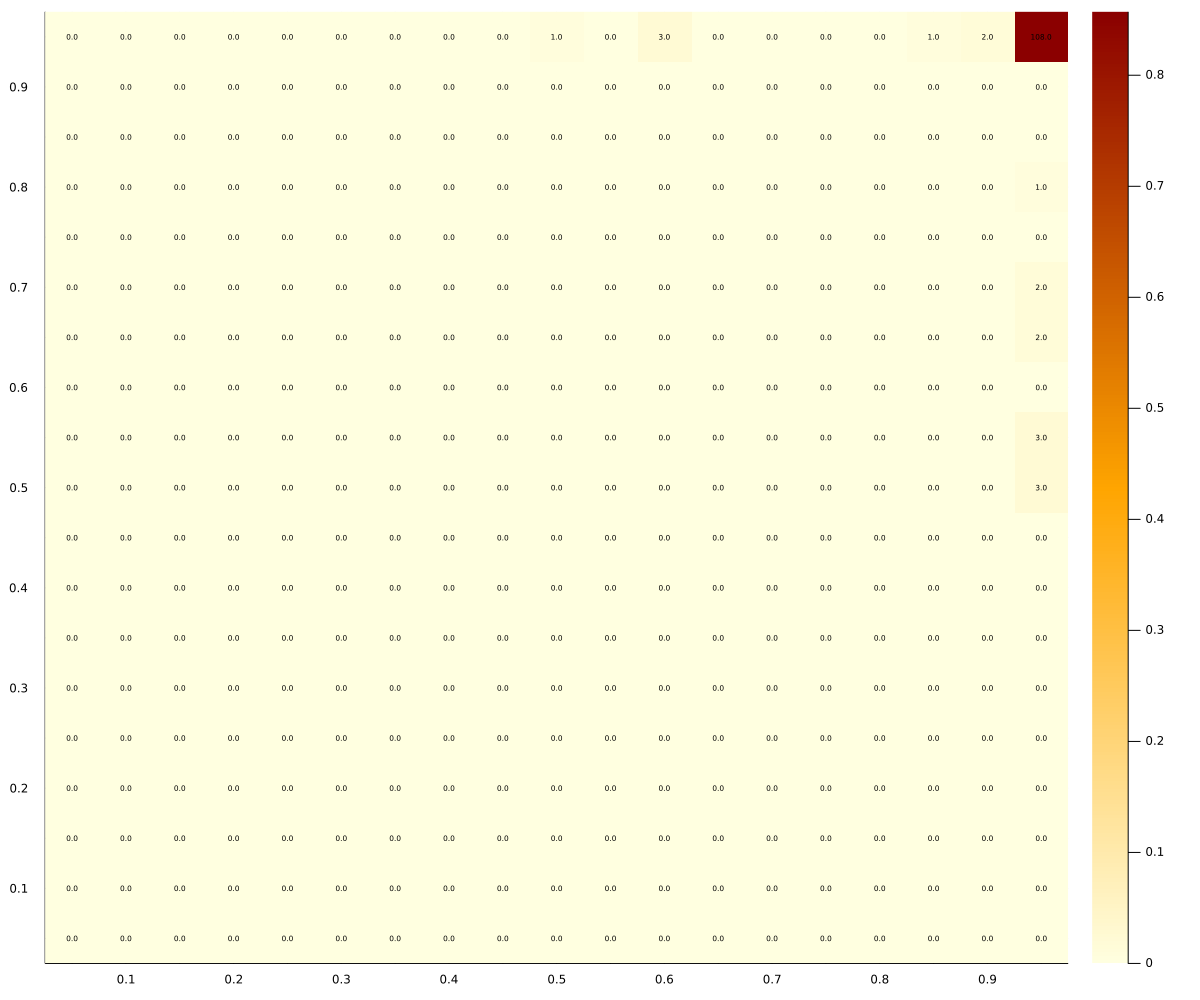}
  \caption{SPA bids, $\gamma = 0.999$}
\end{subfigure}
\caption{Outcomes of 500 simulations for the first two players in a three bidder auction.}
\label{fig:three}
\end{figure}

\end{appendices}

\end{document}